\documentclass[twocolumn,aps,pra,amssymb,amstext,showpacs,longbibliography]{revtex4-1}

\usepackage{amsfonts,amssymb,amsmath,mathrsfs,graphicx,color}

\newcommand{\real}{\mathrm{Re}}

\newcommand{\afm}{\mathrm{AFM}}
\newcommand{\dpm}{\mathrm{DPM}}

\newcommand{\ket}[1]{|#1\rangle}

\begin{document}

\title{Comparison between two models of absorption of matter waves by
  a thin time-dependent barrier}

\author{Maximilien Barbier$^1$, Mathieu Beau$^{2,3}$, Arseni Goussev$^1$}

\affiliation{$^1$Department of Mathematics and Information Sciences,
  Northumbria University, Newcastle Upon Tyne, NE1 8ST, United
  Kingdom\\ $^2$Department of Physics, University of Massachusetts,
  Boston, MA 02125, USA \\$^3$Dublin Institute for Advanced Studies,
  School of Theoretical Physics, 10 Burlington Road, Dublin 4,
  Ireland}

\date{\today}

\begin{abstract}
  We report a quantitative, analytical and numerical, comparison
  between two models of the interaction of a non-relativistic quantum
  particle with a thin time-dependent absorbing barrier. The first
  model represents the barrier by a set of time-dependent
  discontinuous matching conditions, which are closely related to
  Kottler boundary conditions used in stationary wave optics as a
  mathematical basis for Kirchhoff diffraction theory. The second
  model mimics the absorbing barrier with an off-diagonal
  $\delta$-potential with a time-dependent amplitude. We show that the
  two models of absorption agree in their predictions in a
  semiclassical regime -- the regime readily accessible in modern
  experiments with ultracold atoms.
\end{abstract}


\maketitle


\section{Introduction}
\label{sec:intro}

The concept of matter-wave absorption often proves valuable in
modeling irreversible escape or leakage of a quantum wave function
from a practically ``interesting'' part of the system's Hilbert space
into the complementary, ``uninteresting'' part. A representative
example is an inelastic atomic (or molecular) collision process, in
which a colliding atom may end up ionized (or a molecule dissociated),
thus becoming invisible for a detector. An analytical or numerical
description of such a process may be significantly simplified by
restricting the full wave function to a subspace of detectable states,
and mimicking its coupling to the complementary subspace of
undetectable (ionized or dissociated) states by endowing the system's
Hamiltonian with such nonhermitian features as absorbing complex
potentials~\cite{MPNE04Complex} or absorbing boundary
conditions~\cite{Had91Transparent,Shi91Absorbing,Kus92Absorbing,PBT12Almost,TZO14Absorption}.

The present study is devoted to a problem of the motion of a
one-dimensional quantum particle in the presence of a point-like
absorber whose absorbing properties change in the course of time. On a
practical side, this problem models for instance a passage of an atom
through a partially ionizing laser light sheet of time-dependent
intensity. On a more fundamental level, the problem is a
generalization of the celebrated Moshinsky shutter
problem~\cite{Mos52Diffraction,Mos76Diffraction,MMS99Diffraction},
which is a paradigm of the theory of quantum
transients~\cite{Kle94Exact,MME08Time,MRC09Time,CGM09Quantum}. As
first shown by Moshinsky, a sudden removal of a completely absorbing
point-like barrier may give rise to well-pronounced oscillations of
the amplitude of the particle's wave function, and the mathematical
structure of these oscillations is closely analogous to that of
intensity fringes observed in optical diffraction of light at
apertures with straight edges. This temporal quantum phenomenon is
commonly referred to as ``diffraction in time''. In addition to its
purely theoretical interest, diffraction in time has been at the heart
of many experimental studies
\cite{SGA+96Atomic,HFG+98Matter,LSW+05Attosecond,CMPL05Diffraction,PB09Double}.

Here we consider a one-dimensional quantum particle characterized by a
wave function $\Psi(x,t)$, with $x$ and $t$ denoting the spatial
coordinate and time respectively. As we will be dealing with an
absorbing system, the norm of the wave function may generally be
smaller than unity, i.e., $\int_{-\infty}^{+\infty} dx \,
|\Psi(x,t)|^2 \leq 1$. We assume however that the initial state
$\Psi_0(x) \equiv \Psi(x,0)$ is normalized to unity,
$\int_{-\infty}^{+\infty} dx \, |\Psi_0(x)|^2 = 1$, and is spatially
localized, for the concreteness, to the left of the origin, so that
$\Psi_0(x) = 0$ for all $x \geq 0$. We further imagine that a
point-like barrier is positioned at $x=0$; the barrier is such that it
partly absorbs matter waves passing through it, and that the
proportion of the amount absorbed to the amount transmitted depends on
time $t$. (Below we will present a more concrete definition of the
absorption process.) The central aim of the present work is to compare
two different approaches that allow one to evaluate the particle's
wave function $\Psi(x,t)$ in the transmission region, $x > 0$, at time
$t > 0$.

The first approach, which we will refer to as the aperture function
model (AFM), was originally devised in
Refs.~\cite{Gou12Huygens,Gou13Diffraction}. It is based on modeling
the absorbing barrier by means of discontinuous time-dependent
boundary conditions at $x = 0$, connecting the values of the wave
function and its spatial derivative across the barrier. The main
strength of the AFM is that it is exactly
solvable~\cite{Gou13Diffraction}. This quality of the model allows for
accurate and essentially analytical evaluation of the part of the wave
function transmitted through the barrier. In particular, the AFM has
been recently used to advance absorption-based displacement,
splitting, squeezing, and cooling of atomic wave packets
\cite{Gou15Manipulating}. An apparent weakness of the AFM however is
the absence of a first-principle derivation of the absorbing boundary
conditions. This is why a careful comparison between the AFM and some
first-principle model of time-dependent absorption is much needed.

The second approach to the problem, termed here the delta potential
model (DPM), is a time-dependent extension of an atom-laser system
studied in Ref.~\cite{CM06Exact}. In this model, the moving particle
is a two-level atom, with the internal states labelled $|1\rangle$ and
$|2\rangle$. The atom is regarded as detectable (visible) if it is in
the internal state $|1\rangle$, and undetectable (invisible or
absorbed) if it is in $|2\rangle$. Initially, the atom is detectable,
and its total state is the product state $\Psi_0(x) |1\rangle$. In the
course of its motion, the atom interacts with a time-dependent
off-diagonal $\delta$-potential, representing an absorbing barrier,
whose role is to mix populations of the internal states $|1\rangle$
and $|2\rangle$. As a result of this interaction, the total state of
the atom at time $t > 0$ will generally be a linear combination of the
two internal states. It is the projection on $|1\rangle$ that
determines the part of the wave function that has not been absorbed by
the barrier. An important advantage of the DPM over the AFM is that
the former has a solid first-principle justification. Its main
disadvantage however is that, except for few special cases, the DPM
does not admit analytical treatment, and the transmitted wave function
has to be computed numerically.

In this paper we make a comparison between the AFM and DPM. We show
that in the semiclassical (short-wavelength) regime the two models
agree in their predictions of the wave function transmitted through
the barrier. In an atom-optics setting, this semiclassical regime is
of special importance as it covers a wide range of experimentally
relevant parameters. Outside the semiclassical regime however we
generally find quantitative discrepancies between the predictions of
the two models.

The paper is organized as follows. In Sec.~\ref{sec:models}, we give a
detailed definition of the two models of absorption, the AFM and DPM,
and hypothesize about a connection between them. A semiclassical
justification of the connection is presented in
Sec.~\ref{sec:semiclassics}. A numerical study of the connection, both
within and outside the semiclassical regime, is reported in
Sec.~\ref{sec:numerics}. Finally, in Sec.~\ref{sec:conclusion}, we
discuss our results and make concluding remarks. Some technical
details are deferred to an appendix.


\section{Statement of the problem}
\label{sec:models}

In this section we define, in full detail, two mathematical models
describing the motion of a non-relativistic quantum particle in the
presence of a time-dependent absorbing barrier, and outline our
strategy for making a comparison between them.


\subsection{Aperture function model}

We start by summarizing the AFM, originally developed in
Refs.~\cite{Gou12Huygens,Gou13Diffraction}. As specified in
Sec.~\ref{sec:intro}, we consider a quantum particle initially
described by a wave function $\Psi_0(x)$, which is assumed to be
entirely localized on the half-line $x<0$. A time-dependent absorbing
barrier is positioned at $x=0$. In a time $t > 0$, $\Psi_0(x)$ evolves
into a wave function $\Psi_{\afm}(x,t)$. Below we define the laws
governing this evolution.

In the AFM model, the wave function $\Psi_{\afm}$ is taken to satisfy
the time-dependent Schr\"{o}dinger equation at all times and
everywhere away from the barrier, i.e.,
\begin{equation}
  \left( i \frac{\partial}{\partial \tau} + \frac{\hbar}{2 m}
  \frac{\partial^2}{\partial x^2} \right) \Psi_{\afm}(x,\tau) = 0
\label{SE}
\end{equation}
for $0 < \tau < t$ and both $x < 0$ and $x > 0$. The relation between
the wave function and its spatial derivative at $x < 0$ to those at $x
> 0$ is given by the conditions
\begin{align}
  &\Psi_{\afm}(x,\tau) \big|_{x=0^-}^{x=0^+} = -\big[ 1 - \chi(\tau)
    \big] \Psi_{\mathrm{free}}(x,\tau) \big|_{x=0}
  \,, \label{BC-1}\\[0.1cm] &\frac{\partial
    \Psi_{\afm}(x,\tau)}{\partial x} \bigg|_{x=0^-}^{x=0^+} = -\big[ 1
    - \chi(\tau) \big] \frac{\partial
    \Psi_{\mathrm{free}}(x,\tau)}{\partial x} \bigg|_{x=0}
  \,, \label{BC-2}
\end{align}
satisfied for $0 < \tau < t$. Here, $\Psi_{\mathrm{free}}(x,\tau)$ is
the result of a free propagation of the initial state $\Psi_0(x)$
through time $\tau$, i.e.
\begin{equation}
  \Psi_{\mathrm{free}}(x,\tau) = \int_{-\infty}^{+\infty} dx'
  K_0(x-x',\tau) \Psi_0(x') \,,
\label{free_particle_wave_packet}
\end{equation}
where
\begin{equation}
  K_0(\xi,\tau) = \sqrt{\frac{m}{2 \pi i \hbar \tau}} \exp \left( i
  \frac{m \xi^2}{2 \hbar \tau} \right)
\label{free-particle_propagator}
\end{equation}
is the free-particle propagator. The real-valued function $\chi(\tau)$
is the aperture function of the barrier, and is defined in the
following way. The values of $\chi$ range between 0 and 1, with 0
representing the situation when the barrier absorbs entirely all
incident matter waves (completely absorbing, nontransparent barrier)
and 1 corresponding to the unobstructed, free-particle motion
(perfectly transparent barrier). More generally, $\chi^2(\tau)$ is
taken to be an instantaneous (pertinent to the barrier at time $\tau$)
transmission probability defined with respect to the initial state
$\Psi_0(x)$. The matching conditions, given by Eqs.~(\ref{BC-1}) and
(\ref{BC-2}), mimic the action of a time-dependent absorbing barrier;
these conditions are a time-dependent quantum-mechanical
generalization of the black-screen boundary conditions originally put
forward by Kottler as a way of providing a solid mathematical basis
for Kirchhoff diffraction theory
\cite{Kot23Zur,Kot65Diffraction,NHL95Diffraction}. Finally, in
addition to the Schr\"{o}dinger equation, Eq.~(\ref{SE}), and the
matching conditions, Eqs.~(\ref{BC-1}) and (\ref{BC-2}), the wave
function is required to satisfy Dirichlet boundary conditions at
infinity,
$\lim\limits_{x \rightarrow \pm \infty} \Psi_{\afm}(x,\tau) = 0$, and
the initial condition $\Psi_{\afm}(x,0) = \Psi_0(x)$.

The main benefit of the AFM, formulated above, is that it has a unique
exact solution, which in the transmission region, for $x > 0$, can be
written as \cite{Gou13Diffraction}
\begin{equation}
  \Psi_{\afm}(x,t) = \int\limits_{-\infty}^0 dx' K_{\afm}(x,x',t)
  \Psi_0(x')
\label{Psi_AFM_general}
\end{equation}
with
\begin{align}
  K_{\afm}(x,x',t) = \frac{1}{2} \int\limits_0^t &d\tau \, \chi(\tau)
  \left( \frac{x}{t-\tau} - \frac{x'}{\tau} \right)
  \nonumber\\ &\times K_0(x,t-\tau) K_0(x',\tau) \,.
\label{K_AFM_general}
\end{align}
The propagator $K_{\afm}$ admits the following physical
interpretation. The amplitude of a particle's passage from a point $x'
< 0$ to a point $x > 0$ in time $t$ is determined by a superposition
of a continuous family of paths parametrized by $\tau$. Each of these
paths consists of (i) a free flight from $x'$ to the barrier in time
$\tau$, (ii) a modulation of the amplitude by a factor proportional to
the aperture function $\chi(\tau)$ and the mean velocity $\frac{1}{2}
\left( \frac{x}{t-\tau} - \frac{x'}{\tau} \right)$ at which the
particle crosses the barrier, and (iii) another free flight from the
barrier to $x$ in the remaining time $t-\tau$.

Finally, we note that the AFM is conceptually similar to other
analytical approaches devised to describe the wave function
transmission through real, reflecting time-dependent barriers (see,
e.g.,
Ref.~\cite{BZ97Diffraction,BEM01Sources,CMM07Time,TMB+11Explanation}). Relation
between quantum propagators for absorbing and reflecting barriers has
been recently discussed in Ref.~\cite{BD15Three}.

\subsection{Delta potential model}

In Ref.~\cite{CM06Exact}, an exact propagator was obtained describing
the motion of an atom interacting with a point-like,
$\delta$-potential laser whose frequency was in resonance with a given
interatomic transition. (The corresponding problem for a laser with a
semi-infinite spatial extent was studied in
Ref.~\cite{NEM+03Suppression}.) Here we formulate an extended version
of the point-like atom-laser interaction model in which the intensity
of the $\delta$-laser is allowed to change in time in accordance with
an externally prescribed protocol.

To this end, we consider a two level atom, with the internal states
labelled $|1\rangle$ and $|2\rangle$. At any time $\tau$, the full
state of the atom can be written as $\psi_1(x,\tau) |1\rangle +
\psi_2(x,\tau) |2\rangle$, with $\psi_1$ and $\psi_2$ representing the
spatial parts of the state. The atom is initially prepared in the
state defined by $\psi_1(x,0) = \Psi_0(x)$ and $\psi_2(x,0) = 0$. In
the course of its time evolution, the atom propagates freely
everywhere in space except for the point $x=0$, at which a barrier is
placed. The barrier is represented by an off-diagonal
$\delta$-potential with a time-dependent amplitude, $V(x,\tau) = \hbar
\Omega(\tau) \delta(x) \big( | 1 \rangle \langle 2 | + | 2 \rangle
\langle 1 | \big)$. The amplitude $\Omega(\tau)$ is a positive-valued
function of time quantifying the strength of the atom-laser coupling
and directly proportional to the square root of the laser intensity. So,
the time evolution of the full atomic state is governed by the
time-dependent Schr\"{o}dinger equation
\begin{equation}
  i \hbar \frac{\partial}{\partial \tau} \left(
  \begin{array}{c}
    \psi_1 \\ \psi_2
  \end{array} \right) = H_{\dpm} \left(
  \begin{array}{c}
    \psi_1 \\ \psi_2
  \end{array} \right)
\label{SE-2}
\end{equation}
for $0 < \tau < t$, where
\begin{equation}
  H_{\dpm} = H_0 + V(x,\tau)
\label{Hamil}
\end{equation}
with
\begin{equation}
  H_0 = -\frac{\hbar^2}{2 m}
    \frac{\partial^2}{\partial x^2} \left(
  \begin{array}{cc}
    1 & 0 \\ 0 & 1
  \end{array} \right)
\label{H0}
\end{equation}
and
\begin{equation}
  V(x,\tau) = \hbar \Omega(\tau) \delta(x) \left(
  \begin{array}{cc}
    0 & 1 \\ 1 & 0
  \end{array} \right) \,.
\label{V(x,t)}
\end{equation}
In addition, the full wave function is subject to the usual Dirichlet
boundary conditions at infinity, $\lim\limits_{x \rightarrow \pm
  \infty} \psi_1(x,\tau) = 0$ and $\lim\limits_{x \rightarrow \pm
  \infty} \psi_2(x,\tau) = 0$.

In the context of the DPM, the internal state $| 1 \rangle$ represents
detectable particles, and $| 2 \rangle$ labels particles absorbed by
the barrier. Thus, the atom that at time $t$ has traversed the barrier
and remained in its original internal state $| 1 \rangle$, or, in
other words, the particle that has survived the absorbing barrier, is
described by the wave function
\begin{equation}
  \Psi_{\dpm}(x,t) = \psi_1(x,t) \,.
\end{equation}
The wave function $\Psi_{\dpm}$ admits analytical evaluation only in
very few special cases. One such case is that of a time-independent
potential, $\Omega(\tau) = \Omega_0$ \cite{CM06Exact}. Another exactly
solvable case corresponds to $\Omega(\tau) = \Omega_1 / \tau$, with
$\Omega_1$ being a constant; here, the exact propagator is obtained by
first making the substitution $\psi_{\pm} = \psi_1 \pm \psi_2$ to
decouple Eq.~(\ref{SE-2}), and then using a known solution for the
problem of a one-dimensional particle in the $\tau^{-1} \delta(x)$
potential \cite{SK88adiabaticity}. However, in general, i.e. for an
arbitrary function $\Omega(\tau)$, Eq.~(\ref{SE-2}) can only be
tackled by means of a direct numerical integration.

\subsection{Connection between the two models}

In this paper we make a quantitative comparison between
$\Psi_{\afm}(x,t)$ and $\Psi_{\dpm}(x,t)$ in the transmission region,
$x > 0$. The initial wave function of the particle is taken to be a
Gaussian wave packet:
\begin{equation}
  \Psi_0(x) = \left( \frac{1}{\pi \sigma^2} \right)^{1/4} \exp \left[
    -\frac{(x-x_0)^2}{2 \sigma^2} + i \frac{m v_0}{\hbar} (x-x_0)
    \right] \,,
\label{Psi0}
\end{equation}
where $m$ is the mass of the particle, $x_0$ and $v_0$ are its average
position and velocity, respectively, and $\sigma$ is the spatial extent
of the wave packet. Hereinafter we consider $x_0 < 0$, $|x_0| \gg
\sigma > 0$, and $v_0 > 0$. The reason for our choice of the initial
wave function is twofold. On the one hand, having $\Psi_0(x)$ given by
a simple Gaussian facilitates analytical treatment of the problem. On
the other hand, localized wave packets with a nonzero average velocity
can be routinely generated in laboratory experiments with ultracold
atoms (see, e.g.,
Refs.~\cite{FCG+11Realization,CFV+13Matter,JMR+12Coherent}).

In order to compare the two models of absorption, the AFM and DPM, we
first need to specify a relation
$\chi(\tau) = \mathcal{T}\big( \Omega(\tau) \big)$ connecting the
aperture function $\chi$ with the atom-laser coupling strength
$\Omega$ at any time $0 < \tau < t$. Here we do this in the following
intuitive way: We take the function $\mathcal{T}(\Omega_0)$ to be the
$|1\rangle$-channel transmission amplitude associated with a plane
wave of momentum $m v_0$ incident upon a $\delta$-barrier of constant
strength $\Omega_0$. More concretely, we take
$\Omega(\tau) = \Omega_0$ and consider a scattering state solution to
Eq.~(\ref{SE-2}) of the form
\begin{align}
  \left(
  \begin{array}{c}
    \psi_1 \\ \psi_2
  \end{array} \right) = e^{i (k x - \omega t)} \left\{
  \begin{array}{ll}
    \left(
    \begin{array}{c}
      1 \\ 0
    \end{array} \right) + \left(
    \begin{array}{c}
      \mathcal{R}_{11} \\ \mathcal{R}_{21}
    \end{array} \right) e^{-2 i k x} & \; \mathrm{if} \: x < 0 \\[0.4cm]
    \left(
    \begin{array}{c}
      \mathcal{T}_{11} \\ \mathcal{T}_{21}
    \end{array} \right) & \; \mathrm{if} \: x > 0
  \end{array} \right.
\end{align}
with $k = m v_0 / \hbar$ and $\omega = m v_0^2 / 2 \hbar$. The
transmission amplitudes, $\mathcal{T}_{11}$ and $\mathcal{T}_{21}$,
and the reflection amplitudes, $\mathcal{R}_{11}$ and
$\mathcal{R}_{21}$, are found in a standard way by requiring the wave
function to be continuous and to have a discontinuous spatial
derivative at $x = 0$. In particular, one can straightforwardly show
that $\mathcal{T}_{11} = 1 / \big[1 + (m \Omega_0 / \hbar k)^2 \big] =
1 / \big[ 1 + (\Omega_0 / v_0)^2 \big]$. We then take the transmission
amplitude $\mathcal{T}_{11}$ as a definition of the function
$\mathcal{T}(\Omega_0)$, so that, for a time-dependent
$\delta$-barrier, we have
\begin{equation}
  \chi(\tau) = \frac{1}{1 + \left[ \Omega(\tau) / v_0 \right]^2}
\label{chi_vs_Omega}
\end{equation}
for $0 < \tau < t$.

Equipped with Eq.~(\ref{chi_vs_Omega}), it is now feasible to compare
the wave functions $\Psi_{\afm}(x,t)$ and $\Psi_{\dpm}(x,t)$ evolved
from the same initial state $\Psi_0(x)$, given by Eq.~(\ref{Psi0}). In
Sec.~\ref{sec:semiclassics}, we make this comparison analytically in a
semiclassical regime, and in the subsequent section,
Sec.~\ref{sec:numerics}, we go beyond the semiclassical regime by
numerically solving Eq.~(\ref{SE-2}) (or, by using exact analytical
results when available) in various experimentally realistic
scenarios. We summarize our results and make concluding remarks in
Sec.~\ref{sec:conclusion}.


\section{Semiclassical regime}
\label{sec:semiclassics}

The aim of this section is to analytically investigate the validity of
Eq.~(\ref{chi_vs_Omega}) as a proposed connection between the AFM and
DPM. Our analytical calculations are performed in a semiclassical
regime defined with respect to the initial wave packet $\Psi_0(x)$,
given by Eq.~(\ref{Psi0}) and characterized by the mean position $x_0
< 0$, mean velocity $v_0 > 0$, and spatial dispersion $\sigma$. The
semiclassical regime is detailed by the conditions
\begin{align}
 \sigma \ll \lvert x_0 \rvert \lesssim v_0 (t-t_c) \ll \frac{m
   \sigma^2 v_0}{\hbar} \, \text{,}
\label{SC_reg_conditions}
\end{align}
where
\begin{align}
  t_c \equiv \frac{\lvert x_0 \rvert}{v_0}
\label{defSemiClassical_time}
\end{align}
denotes the time needed for the corresponding classical particle to
reach the barrier. The first condition in
Eq.~(\ref{SC_reg_conditions}), $\sigma \ll |x_0|$, specifies that the
initial wave packet is well localized around $x_0$, and allows us to
effectively restrict the support of $\Psi_0(x)$ to the half-line
$x<0$.  The other two conditions, $|x_0| \lesssim v_0 (t-t_c)$ and
$v_0 (t-t_c) \ll m \sigma^2 v_0 / \hbar$, ensure that at time $t$ and
in the absence of a barrier the wave packet would be localized well
inside the transmission region, $x>0$. Indeed, the two conditions
imply that $\hbar t / m \sigma^2 \ll 1$, which means that the spatial
spreading of the wave packet dictated by the Uncertainty Principle is
negligible during the time $t$; this regime is closely related to the
so-called frozen Gaussian approximation \cite{Hel81Frozen}.

We now want to compare the predictions of the AFM and DPM in the
semiclassical regime defined by Eq.~(\ref{SC_reg_conditions}). The
wave function $\Psi_{\afm}(x,t)$ is related to the initial state
$\Psi_{0}(x)$ through Eqs.~(\ref{Psi_AFM_general}) and
(\ref{K_AFM_general}). Similarly, in the DPM, the state $\psi_{1}(x,t)
\ket{1} + \psi_{2}(x,t) \ket{2}$ of the two level atom is related to
its initial state $\Psi_{0}(x) \ket{1}$ through
\begin{align}
  \begin{pmatrix}
    \psi_{1}(x,t) \\
    \psi_{2}(x,t)
  \end{pmatrix} = \int_{- \infty}^{0} dx' \, \widehat{K}(x,x',t)
  \begin{pmatrix}
    \Psi_{0}(x') \\ 0
  \end{pmatrix} \,.
\label{DPM_WF_def}
\end{align}
Here $\widehat{K}$ is a matrix propagator,
\begin{align}
  \widehat{K}(x,x',t) = \begin{pmatrix} K_{11}(x,x',t) &
    K_{12}(x,x',t) \\ K_{21}(x,x',t) & K_{22}(x,x',t)
  \end{pmatrix} \,,
\label{DPM_propa_matrix}
\end{align}
whose first component determines the wave function $\Psi_{\dpm}(x,t)$:
\begin{align}
  \Psi_{\dpm}(x,t) = \int_{- \infty}^{0} dx' \, K_{11}(x,x',t)
  \Psi_{0}(x') \,.
\label{Psi_DPM_def}
\end{align}

In this section, we analytically compare the wave functions
$\Psi_{\afm}(x,t)$ and $\Psi_{\dpm}(x,t)$ in two different
scenarios. First, in Sec.~\ref{slowBarrier_subSec}, we address the
case of a slowly varying time-dependent barrier, and, within the
semiclassical approximation, find an explicit expression for the
propagator $K_{11}(x,x',t)$. Using this expression, we show that if
$\chi(\tau)$ and $\Omega(\tau)$ are related to each other through
Eq.~(\ref{chi_vs_Omega}), the agreement between the two wave
functions, $\Psi_{\afm}(x,t)$ and $\Psi_{\dpm}(x,t)$, is strong in a
spatial region around the center of the corresponding free-particle
wave packet, i.e., around the point $x = x_t$, where the function
$x_{\tau}$ is defined as
\begin{equation}
  x_{\tau} \equiv x_0 + v_0 \tau \,.
\label{defCenter_psi_free}
\end{equation}
Then, in Sec.~\ref{theory:instant_barrier}, we analyze the case of a
rapidly (in fact, instantaneously) varying barrier, and again
demonstrate a strong agreement between the two wave functions in a
neighborhood of the point $x_t$. (We quantify the spatial extent of
the neighborhood in Sec.~\ref{sec:numerics}.)


\subsection{Slowly varying barrier}
\label{slowBarrier_subSec}

Let us first recall that the full propagator $\widehat{K}(x,x',\tau)$
must obey the time-dependent Schr\"{o}dinger equation,
Eq.~(\ref{SE-2}). Therefore, the element $K_{11}(x,x',\tau)$ satisfies
the free-particle time-dependent Schr\"{o}dinger equation on both
sides of the barrier, i.e.
\begin{align}
  \left( \frac{\partial ^2}{\partial x ^2} + \frac{i}{\alpha}
  \frac{\partial}{\partial \tau} \right) K_{11}(x,x',\tau) = 0 \quad
  \mathrm{for} \quad x,x' \neq 0 \,,
\label{TDSE_K_11}
\end{align}
where $\alpha = \hbar /2m$. By definition of a quantum propagator,
$K_{11}(x,x',\tau)$ is subject to the initial condition
\begin{align}
  K_{11}(x,x',0^+) = \delta(x - x') \,.
\label{IC}
\end{align}
Also, Dirichlet boundary conditions imposed on the wave function at
infinity require
\begin{align}
  K_{11}(x \to \pm \infty,x',\tau) = 0 \quad \mathrm{for} \quad \alpha
  = -i \lvert \alpha \rvert \,.
\label{BC_infty}
\end{align}
In addition to Eqs.~\eqref{IC} and~\eqref{BC_infty}, we also know one
matching condition at $x=0$. Indeed, the spatial continuity of the
wave function implies that $K_{11}(x,x',\tau)$ is continuous at $x=0$,
so that
\begin{align}
  K_{11}(x,x',\tau) \big\rvert_{x=0^-}^{x=0^+} = 0 \,.
\label{BC_conti_K_11}
\end{align}
Therefore, we only lack one additional matching condition at $x=0$ in
order to have a well-posed mathematical problem, uniquely determining
$K_{11}(x,x',\tau)$.

Our strategy is as follows. First, within the semiclassical regime and
under the assumption of a slowly varying barrier, we find the missing
matching condition satisfied by the spatial derivative of $K_{11}$ at
$x=0$. Second, we solve the resulting mathematical problem for the
propagator, obtaining an explicit expression for
$K_{11}(x,x',\tau)$. Third, using the expression for the propagator we
make a direct comparison between the wave functions $\Psi_{\afm}(x,t)$
and $\Psi_{\dpm}(x,t)$, finding the two in good agreement.

\subsubsection{Matching condition for $\partial_x K_{11}(x,x',\tau)$ at $x=0$}
\label{BC_subSubSec}

The full propagator $\widehat{K}(x,x',\tau)$ satisfies the
time-dependent Schr\"{o}dinger equation $i \hbar
\frac{\partial}{\partial \tau} \widehat{K} = H_{\dpm} \widehat{K}$
with $H_{\dpm} = H_0 + V(x,\tau)$,
cf. Eqs.~(\ref{SE-2})--(\ref{V(x,t)}). Denoting the matrix
free-particle propagator, corresponding to $H_0$, by
\begin{align}
  \widehat{K}_0(x-x',\tau) \equiv K_0(x-x',\tau)
  \begin{pmatrix} 1 & 0
    \\ 0 & 1
  \end{pmatrix} \,,
\label{2DFreeProp}
\end{align}
we use the time-dependent Lippmann-Schwinger equation \cite{[{See,
    e.g., }][{}]Sch81Techniques} to represent $\widehat{K}(x,x',\tau)$
as a Dyson series:
\begin{widetext}
\begin{align}
  \widehat{K}(x,x',\tau) &= \widehat{K}_0(x-x',\tau) - \frac{i}{\hbar}
  \int_{0}^{\tau} d \tau_1 \int_{-\infty}^{+\infty} d x''
  \widehat{K}_0(x-x'',\tau-\tau_1) V(x'',\tau_1)
  \widehat{K}(x'',x',\tau_1) \nonumber \\ &= \widehat{K}_0(x-x',\tau)
  - i \int_{0}^{\tau} d \tau_1 \widehat{K}_0(x,\tau-\tau_1)
  \Omega(\tau_1)
  \begin{pmatrix}
    0 & 1 \\ 1 & 0
  \end{pmatrix} \widehat{K}(0,x',\tau_1) \nonumber \\
  &= \widehat{K}_0(x-x',\tau) + \sum_{n=1}^{+ \infty}
  \widehat{K}^{(n)}(x,x',\tau) \,,
\label{LS}
\end{align}
where
\begin{multline}
  \widehat{K}^{(n)}(x,x',\tau) = (- i)^n \int_{0}^{\tau} d \tau_n
  \int_{0}^{\tau_n} d \tau_{n-1} \ldots \int_{0}^{\tau_2} d \tau_1 \,
  K_0(x,\tau-\tau_n) \Omega(\tau_n) \\ \times K_0(0,\tau_n-\tau_{n-1})
  \ldots \Omega(\tau_2) K_0(0,\tau_2-\tau_1) \Omega(\tau_1)
  K_0(x',\tau_1)
  \begin{pmatrix}
    0 & 1 \\ 1 & 0
  \end{pmatrix} ^n \,.
\label{nOrderTermProp}
\end{multline}
We can readily see from Eq.~\eqref{nOrderTermProp} that the element
$K_{11}^{(n)}(x,x',\tau) \equiv \left( \widehat{K}^{(n)}(x,x',\tau)
\right)_{11}$ vanishes for odd $n$,
\begin{align}
  K_{11}^{(n)}(x,x',\tau) = 0 \quad \mathrm{for} \quad n = 2k + 1
  \quad \mathrm{with} \quad k \geq 0 \,,
\label{OddTermProp}
\end{align}
and, for even $n$, is given by
\begin{multline}
  K_{11}^{(n)}(x,x',\tau) = (- i)^{n} \int_{0}^{\tau} d \tau_{n}
  \int_{0}^{\tau_{n}} d \tau_{n-1} \ldots \int_{0}^{\tau_2} d \tau_1
  \, K_0(x,\tau-\tau_{n}) \Omega(\tau_{n}) \\ \times
  K_0(0,\tau_{n}-\tau_{n-1}) \ldots \Omega(\tau_2)
  K_0(0,\tau_2-\tau_1) \Omega(\tau_1) K_0(x',\tau_1) \quad
  \mathrm{for} \quad n = 2k \quad \mathrm{with} \quad k \geq 1 \,.
\label{EvenTermProp}
\end{multline}
Therefore, in the rest of this section, we treat $n$ as an even
integer, i.e, $n = 2 k$ with $k \ge 1$.

Let us now calculate the partial derivative of
$K_{11}^{(n)}(x,x',\tau)$ with respect to $x$. To this end, we first
rewrite Eq.~\eqref{EvenTermProp} as
\begin{equation}
  K_{11}^{(n)}(\xi_{n+1},\xi_0,\tau_{n+1}) = (-i)^{n}
  \lim\limits_{\xi_n \to 0 } \ldots \lim\limits_{\xi_1 \to 0 }
  \int_{0}^{\tau_{n+1}} d \tau_{n} \ldots \int_{0}^{\tau_2} d \tau_1
  \left( \prod_{j=1}^{n} \Omega(\tau_j) \right) \left( \prod_{j=0}^{n}
  K_0(\xi_{j+1}-\xi_{j},\tau_{j+1}-\tau_{j}) \right) \,,
\label{PropEps}
\end{equation}
\end{widetext}
where, in order to simplify the subsequent calculations, we have
introduced $\xi_{n+1} \equiv x$, $\xi_{0} \equiv x'$, $\tau_{n+1}
\equiv \tau$ and $\tau_0 \equiv 0$. Then, differentiating
Eq.~\eqref{PropEps} with respect to $x = \xi_{n+1}$, and introducing
the dimensionless parameters $\epsilon_{j} \equiv \xi_j/ \lvert x_0
\rvert$ and $\eta_{j} \equiv \tau_j/\tau \equiv \tau_j/\tau_{n+1}$, we
get, in view of Eq.~(\ref{free-particle_propagator}),
\begin{align}
  \frac{\partial}{\partial x} &K_{11}^{(n)}(x,x',\tau) \nonumber \\ &=
  \frac{1}{\lvert x_0 \rvert}\frac{\partial}{\partial \epsilon_{n+1}}
  K_{11}^{(n)}( \lvert x_0 \rvert \epsilon_{n+1},\lvert x_0 \rvert
  \epsilon_0,\tau_{n+1} \eta_{n+1}) \nonumber \\ &=
  \lim\limits_{\epsilon_n \to 0 } \ldots \lim\limits_{\epsilon_1 \to
    0} I^{(n)}(\epsilon_0, \ldots, \epsilon_{n+1},\eta_{n+1})
\label{LimPropExpr}
\end{align}
with
\begin{multline}
  I^{(n)}(\epsilon_0, \ldots, \epsilon_{n+1},\eta_{n+1}) =
  \int_{0}^{\eta_{n+1}} d \eta_{n} \ldots \int_{0}^{\eta_2} d \eta_1
  \\ \times F(\eta_1, \ldots , \eta_n) e^{i \lambda \phi(\eta_1,
    \ldots , \eta_n)} \,,
\label{DefI_n}
\end{multline}
where $\lambda = m x_0^2/2 \hbar \tau$, the amplitude $F$ is defined as
\begin{multline}
  F(\eta_1, \ldots , \eta_n) = (- i)^{n} \left( \frac{m}{2 i \pi
    \hbar} \right)^{\frac{n+1}{2}} \left( \prod_{j=1}^{n} \Omega(\tau
  \eta_j) \right) \\ \times \frac{i m \lvert x_0 \rvert
    (\epsilon_{n+1}-\epsilon_n)}{\hbar (\eta_{n+1} - \eta_{n})}
  \frac{\tau^{\frac{n-3}{2}}}{\sqrt{\prod\limits_{j=0}^{n}
      (\eta_{j+1}-\eta_{j})}} \, \text{,}
\label{DefF}
\end{multline}
and the phase $\phi$ as
\begin{equation}
  \phi(\eta_1, \ldots , \eta_n) = \sum_{j=0}^{n} \frac{(\epsilon_{j+1}
    - \epsilon_j)^2}{\eta_{j+1}-\eta_j} \,.
\label{DefPhase}
\end{equation}

We now compute $I^{(n)}(\epsilon_0, \ldots,
\epsilon_{n+1},\eta_{n+1})$ in the semiclassical regime, defined by
Eq.~(\ref{SC_reg_conditions}). Keeping in mind that $\lambda \gg 1$ in
the semiclassical regime, we evaluate the $n$-dimensional integral in
Eq.~\eqref{DefI_n} by using the stationary phase approximation
\cite{Sch81Techniques}.

First, we find the stationary point $\boldsymbol{\eta}^{(s)} = \left(
\eta_1^{(s)}, \ldots , \eta_n^{(s)} \right)$, defined by the system of
equations
\begin{equation}
  \left. \frac{\partial \phi}{\partial \eta_j} \right\rvert_{\left(
    \eta_1, \ldots , \eta_n \right) \, = \, \boldsymbol{\eta} ^{(s)}}
  = 0 \,,
\label{DefStatPoint}
\end{equation}
for all $1 \leq j \leq n$, and the constraint
\begin{equation}
  0 < \eta_{1}^{(s)} < \ldots < \eta_{n}^{(s)} < 1 \,.
\label{StatPointCond}
\end{equation}
The condition given by Eq.~\eqref{StatPointCond} is needed to ensure
the existence of a neighbourhood of the stationary point
$\boldsymbol{\eta} ^{(s)}$ that is entirely contained inside the
integration domain of Eq.~\eqref{DefI_n}. It can be straightforwardly
verified that the unique solution to Eqs.~(\ref{DefStatPoint}) and
(\ref{StatPointCond}) is
\begin{equation}
  \eta_{j}^{(s)} = \frac{\sum\limits_{k=1}^{j} \lvert \epsilon_{k} -
    \epsilon_{k-1} \rvert}{\sum\limits_{k=1}^{n+1} \lvert \epsilon_{k}
    - \epsilon_{k-1} \rvert}
\label{StatPointExpr}
\end{equation}
for all $1 \leq j \leq n$.

The stationary phase evaluation of the integral in Eq.~(\ref{DefI_n})
proceeds in the standard way. We restrict the integration domain to a
neighbourhood of the stationary point $\boldsymbol{\eta}^{(s)}$ and
replace the phase $\phi(\eta_1, \ldots , \eta_n)$ by its second order
Taylor expansion in powers of $\left( \eta_j - \eta_j^{(s)}
\right)$. Then, assuming that $\Omega(\tau)$ is a slowly, essentially
algebraically varying function of time, we replace the function
$F(\eta_1, \ldots , \eta_n)$ by its value at the stationary point, $F
\left( \eta_1^{(s)}, \ldots , \eta_n^{(s)} \right)$, and take it
outside the $n$-dimensional integral in Eq.~(\ref{DefI_n}). Finally,
extending the integration region to $\mathbb{R}^n$ and performing the
$n$-dimensional Gaussian integration \cite{[{See, e.g.,
  }][{}]Zin10Path}, we obtain
\begin{equation}
  I^{(n)}(\epsilon_0, \ldots, \epsilon_{n+1},\eta_{n+1}) = \left(
  \frac{2 \pi}{\lambda} \right)^{\frac{n}{2}} \frac{F^{(s)} e^{i
      \lambda \phi^{(s)} + i n \pi / 4}}{\sqrt{\det(\mathcal{H})}} \,.
\label{nDGaussianI_n}
\end{equation}
Here, $F^{(s)} \equiv F \left( \eta_1^{(s)}, \ldots , \eta_n^{(s)}
\right)$ and $\phi^{(s)} \equiv \phi \left( \eta_1^{(s)}, \ldots ,
\eta_n^{(s)} \right)$ are the values of the amplitude and phase at the
stationary point, respectively, and $\mathcal{H}$ is the $n$-by-$n$
Hessian matrix, with elements defined as
\begin{align}
  \mathcal{H}_{jk} = \left. \frac{\partial ^{2} \phi}{\partial \eta_j
    \, \partial \eta_k} \right\rvert_{\left( \eta_1, \ldots , \eta_n
    \right) \, = \, \boldsymbol{\eta} ^{(s)}} \,,
\label{DefHessian}
\end{align}
where $1 \le j \le n$ and $1 \le k \le n$.

Substituting Eq.~\eqref{DefPhase} into Eq.~(\ref{DefHessian}), we see
that the Hessian matrix is symmetric and tridiagonal, and its elements
are given by
\begin{subequations}
\label{HessianElem}
\begin{equation}
  \mathcal{H}_{jj} = 2 \left( \sum\limits_{k=1}^{n+1} \lvert
  \epsilon_{k} - \epsilon_{k-1} \rvert \right)^3 \left(
  \frac{1}{\lvert \epsilon_{j} - \epsilon_{j-1} \rvert} +
  \frac{1}{\lvert \epsilon_{j+1} - \epsilon_{j} \rvert} \right)
\label{HDiag}
\end{equation}
for all $1 \leq j \leq n$,
\begin{equation}
  \mathcal{H}_{j \, , \, j+1} = \mathcal{H}_{j+1 \, , \, j} = - 2
  \frac{\left( \sum\limits_{k=1}^{n+1} \lvert \epsilon_{k} -
    \epsilon_{k-1} \rvert \right) ^3}{\lvert \epsilon_{j+1} -
    \epsilon_{j} \rvert}
\label{HNoDiag}
\end{equation}
\end{subequations}
for all $1 \leq j \leq n-1$, and $H_{jk} = 0$ for all $|j-k| \ge
2$. The determinant of the Hessian matrix is given by (see
Appendix~\ref{appA} for details of the calculation)
\begin{equation}
  \det(\mathcal{H}) = 2^n \frac{\left( \sum\limits_{k=1}^{n+1} \lvert
    \epsilon_{k} - \epsilon_{k-1} \rvert
    \right)^{3n+1}}{\prod\limits_{k=1}^{n+1} \lvert \epsilon_{k} -
    \epsilon_{k-1} \rvert} \,.
\label{detHessian}
\end{equation}
We then use Eqs.~\eqref{DefF}, \eqref{DefPhase},
\eqref{StatPointExpr}, and \eqref{detHessian} in
Eq.~\eqref{nDGaussianI_n}, substitute the resulting expression for
$I^{(n)}(\epsilon_0, \ldots, \epsilon_{n+1},\eta_{n+1})$ into
Eq.~\eqref{LimPropExpr}, and take the limits $\epsilon_1 \to 0$,
$\ldots$, $\epsilon_n \to 0$ to obtain
\begin{multline}
  \frac{\partial}{\partial x} K_{11}^{(2k)}(x,x',\tau) =
  \mathrm{sgn}(x) \frac{m \left( \lvert x \rvert - x' \right)}{i \hbar
    \tau} (-1)^{k+1} \\ \times K_0(\lvert x \rvert - x',\tau) \left [
    \frac{\tau^2}{(\lvert x \rvert - x')^2} \Omega \left( \frac{x'}{x'
      - \lvert x \rvert} \, \tau \right) ^2 \right]^k
\label{PropaDerFinal}
\end{multline}
for all $k \ge 1$. Here $\mathrm{sgn}(x) = x / |x|$ denotes the sign
function.

Differentiating Eq.~\eqref{LS} with respect to $x$, and combining the
resulting expression with Eqs.~\eqref{OddTermProp}, and
\eqref{PropaDerFinal}, we find
\begin{align}
  &\frac{\partial}{\partial x} K_{11}(x,x',\tau) = \frac{\partial
    K_0}{\partial x} + \sum_{k=1}^{+ \infty} \frac{\partial}{\partial
    x} K_{11}^{(2k)}(x,x',\tau) \nonumber \\ &= \frac{\partial
    K_0}{\partial x} + \mathrm{sgn}(x) \frac{m (\lvert x \rvert -
    x')}{i \hbar \tau} K_0(\lvert x \rvert - x',\tau)
  \nonumber\\ &\quad \times \sum_{k=1}^{+ \infty} (-1)^{k+1} \left [
    \frac{\tau^2}{(\lvert x \rvert - x')^2} \Omega \left( \frac{x'}{x'
      - \lvert x \rvert} \, \tau \right)^2 \right]^k \,.
\label{DerivPropaSeriesExpr}
\end{align}
Then, Eq.~(\ref{DerivPropaSeriesExpr}) readily gives us the jump of the
spatial derivative at the origin:
\begin{align}
  \frac{\partial}{\partial x} &K_{11}(x,x',\tau)
  \bigg\rvert_{x=0^-}^{x=0^+} \nonumber \\ &= -2 \frac{m x'}{i \hbar
    \tau} K_0(x',\tau) \sum_{k=1}^{+ \infty} (-1)^{k+1} \left [
    \frac{\tau^2}{x'^2} \, \Omega(\tau)^2 \right]^k \,.
\label{JumpCondExpr}
\end{align}
Finally, using the identities $K_0(x',\tau) m x'/i \hbar \tau =
\left. \partial_x K_0(x- x',\tau) \right\rvert_{x=0}$ and
$\sum_{k=1}^{+ \infty} (-1)^{k+1} z^k = z / (1+z)$ for any $z > 0$, we
rewrite Eq.~(\ref{JumpCondExpr}) as
\begin{equation}
  \left. \frac{\partial}{\partial x} K_{11}(x,x',\tau)
  \right\rvert_{x=0^-}^{x=0^+} = - \kappa(x',\tau)
  \left. \frac{\partial}{\partial x} K_0(x-x',\tau) \right\rvert_{x=0}
\label{JumpCondFinal}
\end{equation}
with
\begin{equation}
  \kappa(x',\tau) = \frac{2}{1 +  \left( \frac{x'}{\tau
        \Omega(\tau)} \right)^2 } \,.
\label{defKappa}
\end{equation}
Equation~(\ref{JumpCondFinal}) constitutes the desired matching
condition for the spatial derivative of the propagator.


\subsubsection{Derivation of $K_{11}(x,x',\tau)$}
\label{ExactPropa_subSubSec}

We now solve the initial-boundary value problem formed by the
time-dependent Schr\"{o}dinger equation, Eq.~\eqref{TDSE_K_11},
initial condition, Eq.~\eqref{IC}, and boundary conditions,
Eqs.~\eqref{BC_infty}, \eqref{BC_conti_K_11}
and~\eqref{JumpCondFinal}. Our method of choice is the method of
Laplace transforms. In the rest of this section, we adopt the
following notation for the Laplace transform of a function $f(\tau)$:
\begin{equation}
  \bar{f}(s) = \mathscr{L} \left[ f(\tau) \right] = \int_{0}^{+
    \infty} d\tau e^{- s \tau} f(\tau) \,.
\end{equation}

Taking the Laplace transform of both sides of Eq.~\eqref{TDSE_K_11},
we obtain
\begin{equation}
  \frac{\partial^2}{\partial x^2} \bar{K}_{11}(x,x',s) + \frac{i
    s}{\alpha} \bar{K}_{11}(x,x',s) = \frac{i}{\alpha} \delta(x-x')
\label{LT_TDSE_K_11_withDelta}
\end{equation}
for $x,x' \neq 0$. The structure of Eq.~\eqref{LT_TDSE_K_11_withDelta}
implies that $\bar{K}_{11}(x,x',s)$ is continuous and
$\frac{\partial}{\partial x} \bar{K}_{11}(x,x',s)$ is discontinuous at
$x=x'$. Therefore, Eq.~\eqref{LT_TDSE_K_11_withDelta} is equivalent to
the homogeneous equation
\begin{align}
  \frac{\partial^2}{\partial x^2} \bar{K}_{11}(x,x',s) + \frac{i
    s}{\alpha} \bar{K}_{11}(x,x',s) = 0
\label{LT_TDSE_K_11_withoutDelta}
\end{align}
for $x,x' \neq 0$ and $x \neq x'$, with the matching conditions
\begin{subequations}
\label{LT_BCs}
\begin{equation}
  \left. \bar{K}_{11}(x,x',s) \right\rvert_{x=x'^-}^{x=x'^+} = 0
\label{LT_BC_conti_x_x'}
\end{equation}
and
\begin{equation}
  \left. \frac{\partial}{\partial x} \bar{K}_{11}(x,x',s)
  \right\rvert_{x=x'^-}^{x=x'^+} = \frac{i}{\alpha} \,.
\label{LT_BC_deriv_x_x'}
\end{equation}
Also, taking the Laplace transform of Eqs.~(\ref{BC_infty}),
(\ref{BC_conti_K_11}), and (\ref{JumpCondFinal}), we obtain,
respectively,
\begin{equation}
  \bar{K}_{11}(x \to \pm \infty,x',s) = 0 \quad \mathrm{for} \quad
  \alpha = -i |\alpha| \,,
\label{LT_BCinfty}
\end{equation}
\begin{equation}
  \left. \bar{K}_{11}(x,x',s) \right\rvert_{x=0^-}^{x=0^+} = 0 \,,
\label{LT_BC_conti_x_0}
\end{equation}
and
\begin{equation}
  \left. \frac{\partial}{\partial x} \bar{K}_{11}(x,x',s)
  \right\rvert_{x=0^-}^{x=0^+} = \bar{Q}(s) \,,
\label{LT_BC_deriv_x_0}
\end{equation}
\end{subequations}
where
\begin{align}
  \bar{Q}(s) &= \mathscr{L} \left[ -\kappa(x',\tau)
    \left. \frac{\partial}{\partial x} K_0(x-x',\tau)
    \right\rvert_{x=0} \, \right] \nonumber \\ &= \frac{i x'}{2
    \alpha} \int_{0}^{+ \infty} d\tau \, \mathrm{e}^{- s \tau}
  \kappa(x',\tau) \frac{K_0(x',\tau)}{\tau} \,.
\label{defQ}
\end{align}
Equations~(\ref{LT_TDSE_K_11_withoutDelta}) and (\ref{LT_BCs})
uniquely specify the function $\bar{K}_{11}(x,x',s)$.

Recalling that we are only interested in the case of $x' < 0$, we
define the following three spatial intervals (with respect to a fixed
value of $x'$): the first interval $\mathcal{R}_1$ is the set of all
points $x$ such that $-\infty < x < x'$, the second interval
$\mathcal{R}_2$ corresponds to $x' < x < 0$, and the third interval
$\mathcal{R}_3$ to $0 < x < +\infty$. The general solution of
Eq.~\eqref{LT_TDSE_K_11_withoutDelta} is given by
\begin{align}
  \bar{K}_{11}(x,x',s) = A_j \, \mathrm{e}^{k_+ x} + B_j \,
  \mathrm{e}^{k_- x} \quad \mathrm{for} \quad x \in \mathcal{R}_j \,,
\label{genForm_LT_K_11_region}
\end{align}
where $A_j = A_j(x',s)$ and $B_j = B_j(x',s)$, with $j=1,2,3$, are
arbitrary complex valued functions, and $k_+$ and $k_-$ are given by
\begin{equation}
  k_{\pm} = \pm \, \mathrm{e}^{-i \frac{\pi}{4}}
  \sqrt{\frac{s}{\alpha}} \,.
\label{def_k_pm}
\end{equation}
Restricting $s$ to the complex plane branch $-\pi < \arg (s) < \pi$,
so that $\real \left( \sqrt{s} \right) > 0$, we obtain from
Eq.~\eqref{LT_BCinfty}:
\begin{align}
  B_1 = A_3 = 0 \,.
\label{B_1_A_3_value}
\end{align}
The four remaining conditions, Eqs.~\eqref{LT_BC_conti_x_x'},
\eqref{LT_BC_deriv_x_x'}, \eqref{LT_BC_conti_x_0}, and
\eqref{LT_BC_deriv_x_0}, lead to the matrix equation
\begin{align}
  \begin{pmatrix}
    - \, \mathrm{e}^{k_+ x'} & \mathrm{e}^{k_+ x'} & \mathrm{e}^{k_- -
      x'} & 0 \\ \, k_+ \mathrm{e}^{k_+ x'} & k_+ \mathrm{e}^{k_+ x'}
    & - k_- \mathrm{e}^{k_- x'} & 0 \\ \vphantom{k_+ \mathrm{e}^{k_+
        x'}}{0} & -1 & -1 & 1 \\ \vphantom{k_+ \mathrm{e}^{k_+ x'}}{0}
    & - k_+ & - k_- & k_-
  \end{pmatrix} \begin{pmatrix}
    \vphantom{k_+ \mathrm{e}^{k_+ x'}}{A_1} \\ \vphantom{k_+
      \mathrm{e}^{k_+ x'}}{A_2} \\ \vphantom{k_+ \mathrm{e}^{k_+
        x'}}{B_2} \\ \vphantom{k_+ \mathrm{e}^{k_+ x'}}{B_3}
  \end{pmatrix} = \begin{pmatrix}
    \vphantom{k_+ \mathrm{e}^{k_+ x'}}{0} \\ \vphantom{k_+
      \mathrm{e}^{k_+ x'}}{\frac{i}{\alpha}} \\ \vphantom{k_+
      \mathrm{e}^{k_+ x'}}{0} \\ \vphantom{k_+ \mathrm{e}^{k_+
        x'}}{\bar{Q}}
  \end{pmatrix} \,.
\label{matrixEqCoef}
\end{align}
The solution of this matrix equation gives the remaining coefficients:
\begin{subequations}
\label{coefExpr}
\begin{equation}
  A_1 = \frac{\mathrm{e}^{- i \frac{\pi}{4}}}{2} \,
  \frac{\mathrm{e}^{- k_+ x'}}{\sqrt{\alpha s}} - \,
  \frac{\mathrm{e}^{i \frac{\pi}{4}}}{2} \sqrt{\frac{\alpha}{s}} \,
  \bar{Q}(s) \, \text{,} \label{coefExpr_A1}
\end{equation}
\begin{equation}
  A_2 = - \, \frac{\mathrm{e}^{i \frac{\pi}{4}}}{2}
  \sqrt{\frac{\alpha}{s}} \, \bar{Q}(s) \,
  \text{,} \label{coefExpr_A2}
\end{equation}
\begin{equation}
  B_2 = \frac{\mathrm{e}^{- i \frac{\pi}{4}}}{2} \,
  \frac{\mathrm{e}^{- k_- x'}}{\sqrt{\alpha s}} \,
  \text{,} \label{coefExpr_B2}
\end{equation}
\begin{equation}
  B_3 = \frac{\mathrm{e}^{- i \frac{\pi}{4}}}{2} \,
  \frac{\mathrm{e}^{- k_- x'}}{\sqrt{\alpha s}} - \frac{\mathrm{e}^{i
      \frac{\pi}{4}}}{2} \sqrt{\frac{\alpha}{s}} \, \bar{Q}(s) \,
  \text{.} \label{coefExpr_B3}
\end{equation}
\end{subequations}

Now we substitute Eqs.~\eqref{B_1_A_3_value} and \eqref{coefExpr} into
Eq.~\eqref{genForm_LT_K_11_region}, and take the inverse Laplace
transform on each of the three intervals, $\mathcal{R}_1$,
$\mathcal{R}_3$, and $\mathcal{R}_3$. Here, we use the fact that
$\mathscr{L}^{-1} \left[ \exp \left( -a \sqrt{s} \right) / \sqrt{s}
  \right] = (\pi \tau)^{-1/2} \exp \left(- a^2 / 4 \tau \right)$ for
$\real(a) \geq 0$ \cite{[{See, e.g., }][{}]EMOT54Tables}, and apply
the convolution theorem in computing the inverse Laplace transform of
terms of the form $\big[ \exp \left( -a \sqrt{s} \right) / \sqrt{s}
  \big] \bar{Q}(s)$ with $\real(a) \geq 0$. This computation shows
that the propagator $K_{11}(x,x',\tau)$ has the same expression in all
three spatial intervals, $\mathcal{R}_1$, $\mathcal{R}_3$, and
$\mathcal{R}_3$, reading
\begin{multline}
  K_{11}(x,x',\tau) = K_{0}(x-x',\tau) \\ + \frac{1}{2}
  \int_{0}^{\tau} d \tau_1 \, \frac{x' \kappa(x',\tau_1)}{\tau_1}
  K_{0}(x, \tau - \tau_1) K_{0}(x',\tau_1) \,.
\label{K_11_expr_final}
\end{multline}

\subsubsection{Comparison between AFM and DPM}
\label{chiRelation_subSubSec}

We are now in a position to compare the wave functions
$\Psi_{\afm}(x,t)$ and $\Psi_{\dpm}(x,t)$. First we note that in view
of Eqs.~(\ref{Psi_AFM_general}) and (\ref{K_AFM_general}) and
Eqs.~(\ref{Psi_DPM_def}) and (\ref{K_11_expr_final}), both wave
functions can be written as
\begin{equation}
  \Psi(x,t) = \int_{0}^{t} d \tau \int_{- \infty}^{0} \!\! dx' F
  K_{0}(x,t - \tau) K_{0}(x',\tau) \Psi_{0}(x') \,,
\label{defPsi_slow}
\end{equation}
where $\Psi_{\afm}$ is obtained by taking $F$ to be
\begin{equation}
  F_{\afm}(x,x',\tau,t) \equiv \frac{1}{2} \, \chi(\tau) \left(
  \frac{x}{t - \tau} - \frac{x'}{\tau} \right)
\label{defF_afm_slow}
\end{equation}
and $\Psi_{\dpm}$ by taking $F$ to be
\begin{multline}
  F_{\dpm}(x,x',\tau,t) \equiv \frac{1}{2} \left[ \frac{x}{t - \tau} -
    \frac{x'}{\tau} \left( 1 - \kappa(x',\tau) \right) \right] \\ =
  \frac{1}{2} \left[ \frac{x}{t - \tau} - \frac{x'}{\tau} \left( 1 -
    \frac{2}{1 + \left( \frac{x'}{\tau \Omega(\tau)} \right)^2}
    \right) \right]\,.
\label{defF_dpm_slow}
\end{multline}
Here, we have used Eq.~(\ref{defKappa}), along with the integral
representation of the free-particle propagator
\begin{equation}
  K_0(x-x',t) = \! \int_0^t \frac{d\tau}{2} \! \left( \frac{x}{t-\tau}
  - \frac{x'}{\tau} \right) \! K_0(x,t-\tau) K_0(x',\tau)
\end{equation}
obtained from Eq.~(\ref{K_AFM_general}) by taking $\chi(\tau) = 1$.

We now show that if the functions $\chi$ and $\Omega$ are related
through Eq.~(\ref{chi_vs_Omega}), then the values of
$\Psi_{\afm}(x,t)$ and $\Psi_{\dpm}(x,t)$ are close in the vicinity of
the point $x = x_t$, which is the center of the corresponding free
particle wave packet. Indeed, within the semiclassical regime
specified by Eq.~(\ref{SC_reg_conditions}) and for the values of $x$
close to $x_t$, the dominant contribution to the double integral in
Eq.~(\ref{defPsi_slow}) comes from a neighbourhood of the spatial
point $x' = x_0$, around which the amplitude of $\Psi_0$ is peaked,
and time $\tau = t_c$, at which the corresponding classical particle
reaches the barrier. Since the barrier, and so the function $F$, is
assumed to change in time slowly compared to the exponential terms
contained in $K_0$ and $\Psi_0$, the function $F(x,x',\tau,t)$ in
Eq.~(\ref{defPsi_slow}) can be effectively replaced by $F(x_t, x_0,
t_c, t)$. In the case of the AFM, we have
\begin{equation}
  F_{\afm}(x_t, x_0, t_c, t) = v_0 \chi(t_c) \,,
\label{F_AFM_at_sp}
\end{equation}
while in the case of the DPM,
\begin{equation}
  F_{\dpm}(x_t, x_0, t_c, t) = \frac{v_0}{1 + \left[ \Omega(t_c) / v_0
      \right]^2}\,.
\label{F_DPM_at_sp}
\end{equation}
It is now clear that, in view of Eq.~(\ref{chi_vs_Omega}),
\begin{equation}
  F_{\afm}(x_t, x_0, t_c, t) = F_{\dpm}(x_t, x_0, t_c, t) \,,
\label{relation_chi_kappa_slow}
\end{equation}
and, consequently, $\Psi_{\afm}(x,t) \simeq \Psi_{\dpm}(x,t)$ around
the point $x = x_t$.


\subsection{Moshinsky shutter}
\label{theory:instant_barrier}

We now address a regime opposite to that of a slowly varying barrier.
We consider the case in which the barrier stays completely closed
during $0 < \tau < t_c$, opens instantaneously at time $\tau = t_c$,
and then remains fully open during $t_c < \tau < t$. In the context of
the AFM, this regime is specified by the aperture function
\begin{align}
  \chi(\tau) = \Theta(\tau - t_c) \,,
\label{defAF_Moshi}
\end{align}
where $\Theta$ denotes the Heaviside step function. In the literature,
such an instantaneously opening barrier is commonly referred to as the
Moshinsky shutter (see Ref.~\cite{Mos52Diffraction} for Moshinsky's
original work). According to Eq.~(\ref{chi_vs_Omega}), the atom-laser
coupling strength $\Omega(\tau)$, corresponding to the aperture
function given by Eq.~(\ref{defAF_Moshi}), is
\begin{equation}
  \Omega(\tau) = \left\{
  \begin{array}{ll}
    +\infty & \,, \quad 0 < \tau < t_c \\[0.2cm]
    0 & \,, \quad t_c < \tau < t
  \end{array} \right. \,.
\label{defPot_Moshi}
\end{equation}
Our aim here is to compare the wave packets $\Psi_{\afm}(x,t)$ and
$\Psi_{\dpm}(x,t)$, specified respectively by Eq.~(\ref{defAF_Moshi})
and Eq.~(\ref{defPot_Moshi}), in the transmission region, $x > 0$.

In the AFM case, the wave function can be written as
\begin{multline}
  \Psi_{\afm} (x,t) = \int_{-\infty}^0 d x'' \, K_0(x-x'',t-t_c)
  \\ \times \int_{-\infty}^0 d x' \, K_0(x''-x',t_c) \Psi_0(x') \,.
\label{Psi_AFM_composition_property}
\end{multline}
The equivalence of this composition-property-type representation of
the wave function and the time-integral representation given by
Eqs.~(\ref{Psi_AFM_general}), (\ref{K_AFM_general}), and
(\ref{defAF_Moshi}) has been established in Ref.~\cite{Gou12Huygens}.
Now, within the semiclassical regime specified by
Eq.~(\ref{SC_reg_conditions}), we first evaluate the integral over
$x'$ and then the other integral over $x''$ to find
\begin{multline}
  \Psi_{\afm}(x,t) \\ = \frac{1}{2} \Psi_{\mathrm{fr}}(x,t) \left[ 1 +
    \mathrm{erf} \left( e^{i 3 \pi/4} \sqrt{\frac{m (x - x_{t}) ^2}{2
        \hbar (t - t_c)}} \, \right) \right] \,,
\label{psiAFM_Moshi_expr}
\end{multline}
where $\mathrm{erf(\cdot)}$ denotes the error function, and
\begin{multline}
  \Psi_{\mathrm{fr}}(\xi,\tau) = \left( \frac{1}{\pi \sigma^2}
  \right)^{1/4} \\ \times \exp \left[ - \frac{(\xi - x_{\tau})^2}{2
      \sigma ^2} + i \frac{m v_0}{\hbar} (\xi - x_{\tau}) + i \frac{m
      v_0^2 \tau}{2 \hbar} \right]
\label{defPsi_free}
\end{multline}
is the frozen Gaussian (or nondispersive) approximation of the free
particle wave packet defined by Eq.~(\ref{free_particle_wave_packet}).

We now compute the wave function $\Psi_{\dpm}(x,t)$ corresponding to
Eq.~(\ref{defPot_Moshi}). First, we express the full matrix
propagator $\widehat{K}$ as a sequence of two constant potential
propagators:
\begin{equation}
  \widehat{K}(x,x',t) = \int_{-\infty}^{+\infty} dx'' \,
  \widehat{K}_0(x-x'',t-t_c) \widehat{K}_{\infty}(x'',x',t_c) \,,
\label{K_DPM_compositionRule}
\end{equation}
where $\widehat{K}_0$ is the matrix free-particle propagator defined
by Eq.~(\ref{2DFreeProp}), and
\begin{equation}
  \widehat{K}_{\infty}(\xi_1,\xi_2,\tau) = \lim_{\Omega_0 \to +
    \infty} \widehat{K}_{\Omega_0}(\xi_1,\xi_2,\tau) \,,
\label{defK_inf}
\end{equation}
where $\widehat{K}_{\Omega_0}$ denotes the propagator corresponding to
a time-independent coupling frequency $\Omega(\tau) = \Omega_0$. An
exact expression for $\widehat{K}_{\Omega_0}$ has been derived in
Ref.~\cite{CM06Exact}, and reads
\begin{align}
  &\widehat{K}_{\Omega_0}(\xi_1,\xi_2,\tau) =
  \widehat{K}_0(\xi_1-\xi_2,\tau) \nonumber \\ &\quad -\frac{m
    \Omega_0}{4 \hbar} \sum_{j = \pm 1} e^{j \frac{m \Omega_0}{\hbar}
    (\lvert \xi_1 \rvert + \lvert \xi_2 \rvert)} \mathrm{e}^{i \frac{m
      \Omega_0^2}{2 \hbar} \tau} \mathrm{erfc}(z_j)
  \begin{pmatrix} j
    & 1 \\ 1 & j
  \end{pmatrix}
\label{defK_Omega0}
\end{align}
with
\begin{align}
  z_j = j \sqrt{i \frac{m \Omega_0^2}{2 \hbar} \tau} +
  \sqrt{\frac{m}{2 i \hbar \tau}} (\lvert \xi_1 \rvert + \lvert \xi_2
  \rvert) \,.
\label{def_z_j}
\end{align}
Here, $\mathrm{erfc}(\cdot) = 1 - \mathrm{erf}(\cdot)$ is the
complementary error function.

We readily see from Eq.~\eqref{def_z_j} that $-3 \pi / 4 <
\mathrm{arg} (z_j) < \pi / 4$ as long as $\xi_1, \xi_2 \neq 0$, and
also that $\lim\limits_{\Omega_0 \to + \infty} \lvert z_j \rvert = +
\infty$. Therefore, using the asymptotic expansion \cite{[{See, e.g.,
  }][{}]GR07Table} $\mathrm{erfc} (z_j) \simeq \exp \left( - z_j^2
\right) / \sqrt{\pi} z_j$ in Eq.~(\ref{defK_Omega0}), substituting the
resulting expression into Eq.~(\ref{defK_inf}), and taking the limit
$\Omega_0 \to + \infty$, we obtain
\begin{equation}
  \widehat{K}_{\infty} (\xi_1,\xi_2,\tau) = \widehat{K}_0
  (\xi_1-\xi_2,\tau) - \widehat{K}_0 (|\xi_1| + |\xi_2|,\tau) \,.
\label{K_inf_expr}
\end{equation}
A substitution of Eq.~(\ref{K_inf_expr}) into
Eq.~(\ref{K_DPM_compositionRule}) yields the following expression for
the DPM propagator:
\begin{align}
  &K_{11}(x,x',t) = K_0(x-x',t) \nonumber \\ &\; -\int_{- \infty}^{+
    \infty} dx'' \, K_0(x-x'',t-t_c) K_0 (|x'|+|x''|,t_c) \,.
\label{K_11_Moshinsky}
\end{align}
The DPM wave function is then obtained by substituting
Eq.~(\ref{K_11_Moshinsky}) into Eq.~(\ref{Psi_DPM_def}) and evaluating
the resulting integrals. So, performing the integration over the
initial position $x'$ and using the semiclassical limit, we get
\begin{align}
  &\Psi_{\dpm}(x,t) = \Psi_{\mathrm{fr}}(x,t) \nonumber \\ &\quad -
  \int_{- \infty}^{+ \infty} dx'' \, K_0(x-x'',t-t_c)
  \Psi_{\mathrm{fr}} (|x''|, t_c) \,.
\label{defPsi_DPM_Moshi}
\end{align}
Finally, calculating the integral in Eq.~(\ref{defPsi_DPM_Moshi}), we
establish the following relation between the DPM and AFM wave
functions:
\begin{equation}
  \Psi_{\dpm}(x,t) = \Psi_{\afm}(x,t) + \Delta \Psi(x,t) \,,
\label{Psi_DPM_Moshinksy_final}
\end{equation}
where, within the semiclassical regime defined by
Eq.~(\ref{SC_reg_conditions}),
\begin{equation}
  \Delta \Psi(x,t) = \left( \frac{1}{\pi \sigma^2} \right)^{1/4} \!
  \sqrt{\frac{\hbar (t - t_c)}{2 \pi m (x + x_t)^2}} \; e^{i \beta}
\label{Delta_Psi}
\end{equation}
with $\beta = m x^2 / 2 \hbar (t - t_c) + m v_0^2 t_c / 2 \hbar - 3
\pi / 4$.

It follows from Eqs.~(\ref{Psi_DPM_Moshinksy_final}) and
(\ref{Delta_Psi}) that the values of $\Psi_{\afm}(x,t)$ and
$\Psi_{\dpm}(x,t)$ are close to each other in the vicinity of the
spatial point $x = x_t$. Indeed, for $x$ close to $x_t$, we have
\begin{align}
  |\Delta \Psi| < \sqrt{\frac{\hbar t}{m x_t^2}} \, |\Psi_{\afm}| \ll
  \sqrt{\frac{\hbar t}{m \sigma^2}} \, |\Psi_{\afm}| \ll |\Psi_{\afm}|
  \,.
\end{align}
Far in the tails however, the two wave functions exhibit different
behavior. Indeed, an asymptotic expansion of the right-hand side in
Eq.~(\ref{psiAFM_Moshi_expr}) shows that $|\Psi_{\afm}|^2 \sim 1/x^2$
as $x \to \pm \infty$, whereas the corresponding expansion of
Eq.~(\ref{Psi_DPM_Moshinksy_final}), with
Eqs.~(\ref{psiAFM_Moshi_expr}) and (\ref{Delta_Psi}) taken into
account, yields $|\Psi_{\dpm}|^2 \sim 1/x^4$ as $x \to \pm \infty$.

\bigskip

The analysis presented in this section only establishes agreement
between the predictions of the AFM and DPM in a narrow spatial region
centered around the point $x = x_t$, the point specifying the location
of the corresponding freely-propagating classical particle at time
$t$. In the next section, Sec.~\ref{sec:numerics}, we strengthen our
statement by numerically demonstrating that the agreement between the
two models holds in a much broader spatial region.


\section{Numerical results}
\label{sec:numerics}

In this section we numerically evaluate the wave functions
$\Psi_{\afm}(x,t)$ and $\Psi_{\dpm}(x,t)$, and investigate the
validity of the relation between $\chi$ and $\Omega$,
Eq.~(\ref{chi_vs_Omega}), in both the semiclassical and deep quantum
regimes. We begin by outlining our strategy.

In our numerical study, we focus on two different measures of
similarity between the wave functions. The first one, the so-called
fidelity, is an overlap between $\Psi_{\afm}$ and $\Psi_{\dpm}$ on a
spatial interval $0 < x_A < x < x_B$ at a fixed time $t>0$. Denoted by
$M$, the fidelity is defined as
\begin{align}
  M(t) \equiv \frac{\left\lvert \int_{x_A}^{x_B} dx \,
    \Psi_{\afm}^{\star}(x,t) \Psi_{\dpm}(x,t) \right\rvert
    ^2}{\mathscr{P}_{\afm}(t) \mathscr{P}_{\dpm}(t)} \,,
\label{defFidelity}
\end{align}
where
\begin{align}
  \mathscr{P}_{\afm , \dpm}(t) \equiv \int_{x_A}^{x_B} dx \,
  \left\lvert \Psi_{\afm , \dpm}(x,t) \right\rvert^2
\label{defProba}
\end{align}
represents the probability of finding the particle inside the interval
$x_A < x < x_B$ at time $t$, as predicted by the AFM or DPM,
respectively. The interval boundaries, $x_A$ and $x_B$, are numerical
parameters, and will be further chosen such that the interval contains
the dominant part of the transmitted wave function. By construction,
the fidelity takes values between 0 and 1. $M(t) = 0$ means that the
two wave functions are mutually orthogonal at time $t$, and so
completely different from one another. On the other hand, $M(t) = 1$
is reached whenever the functional form of $\Psi_{\afm}$ is identical
to that of $\Psi_{\dpm}$ up to an arbitrary normalization constant.

Since the fidelity, as defined by Eq.~(\ref{defFidelity}), is
insensitive to the global amplitudes of $\Psi_{\afm}$ and
$\Psi_{\dpm}$, we use the probability ratio
\begin{align}
  R(t) \equiv \frac{\mathscr{P}_{\dpm}(t)}{\mathscr{P}_{\afm}(t)}
\label{defRatio}
\end{align}
as our second tool to compare the AFM and DPM wave functions. It is
worth nothing that $R$ merely compares the overall probabilities of
finding the particle inside the region $x_A < x < x_B$ at time $t$ as
predicted by the $\afm$ and the $\dpm$, and thus complements the
fidelity test.

In this section we numerically evaluate the fidelity, $M(t)$, and
probability ratio, $R(t)$, in four atom-barrier systems that differ
from each other only by the mass of the atom. More specifically, we
consider the dynamics of the alkali atoms $^7 \mathrm{Li}$, $^{23}
\mathrm{Na}$, $^{41} \mathrm{K}$, and $^{87} \mathrm{Rb}$, which are
routinely used in modern ultracold atom-optics experiments. The atomic
masses are $m_{\mathrm{Li}} = 7.016003$~u, $m_{\mathrm{Na}} =
22.989767$~u, $m_{\mathrm{K}} = 40.961825$~u, and $m_{\mathrm{Rb}} =
86.9091805$~u, respectively. Other parameters are the same for all
four systems, and have the following values. The initial wave packet,
defined by Eq.~(\ref{Psi0}), is characterized by the initial position
(with respect to the position of the barrier) $x_0 = -0.15$~mm,
spatial dispersion $\sigma = 30$~$\mu$m, and average velocity $v_0 =
3$~mm/s.  The total propagation time is taken to be $t =
100$~ms. These parameter values imply the classical barrier crossing
time $t_c = 50$~ms~$= t/2$ and the final position of the unperturbed
classical particle $x_t = 0.15$~mm~$= |x_0|$. We note that the chosen
parameter values are comparable to values in real laboratory
experiments \cite{FCG+11Realization,CFV+13Matter,JMR+12Coherent}.

It can be easily seen that while the parameters of the heaviest
(rubidium) system satisfy the semiclassical regime conditions, given
by Eq.~(\ref{SC_reg_conditions}), the parameters of the lightest
(lithium) system do not. Indeed, in the case of $^{87} \mathrm{Rb}$,
we have $m_{\mathrm{Rb}} \sigma^2 v_0 / \hbar \simeq 3.7$~mm, which is
more than 20 times larger than $v_0 (t-t_c) = 0.15$~mm; on the other
hand, in the case of $^7 \mathrm{Li}$, we have $m_{\mathrm{Li}}
\sigma^2 v_0 / \hbar \simeq 0.3$~mm, which is comparable to $v_0
(t-t_c) = 0.15$~mm. Thus, by decreasing the mass of the moving
particle from $m_{\mathrm{Rb}}$ to $m_{\mathrm{Li}}$ we can test the
agreement between the AFM and DPM both within and outside the
semiclassical regime.

Now, having specified the numerical values of all system parameters,
and equipped with Eq.~(\ref{chi_vs_Omega}), we evaluate and compare
the wave functions $\Psi_{\afm}$ and $\Psi_{\dpm}$ in four different
scenarios. First, we consider the case of a time-independent barrier
given by $\Omega(\tau) = \Omega_0$ and
$\chi(\tau) = \chi_0 = \big[ 1 + (\Omega_0/v_0)^2 \big]^{-1}$. Second,
we address the exactly solvable case of a slowly (algebraically)
varying barrier characterized by $\Omega(\tau) = \Omega_1 /
\tau$.
Third, we address barriers whose transparency changes exponentially in
time. And finally, we take a closer look at the instantaneous shutter
case, previously discussed in Sec.~\ref{theory:instant_barrier}.

\subsection{Time-independent barrier}
\label{TimeIndBarSubsec}

We begin by considering the simplest scenario in which $\chi(\tau) =
\chi_0$ and $\Omega(\tau) = \Omega_0$, where $\chi_0$ and $\Omega_0$
are constants related to each other by Eq.~(\ref{chi_vs_Omega}), i.e.,
$\chi_0 = \big[ 1 + (\Omega_0/v_0)^2 \big]^{-1}$. In this case, the
AFM wave function is simply an attenuated free-particle Gaussian wave
packet, $\Psi_{\afm} = \chi_0 \Psi_{\mathrm{free}}$, with
$\Psi_{\mathrm{free}}$ defined by
Eq.~(\ref{free_particle_wave_packet}). The DPM wave function can be
obtained from Eq.~(\ref{Psi_DPM_def}) by taking $K_{11} = \big(
\widehat{K}_{\Omega_0} \big)_{11}$, where $\widehat{K}_{\Omega_0}$ is
the exact DPM propagator given by Eq.~(\ref{defK_Omega0}).

Our definitions of the fidelity and probability ratio, given by
Eqs.~(\ref{defFidelity}) and (\ref{defRatio}), respectively, depend on
the integration region $x_A < x < x_B$. Here, we choose $x_A = x_t - 4
(\Delta x)_t$ and $x_B = x_t + 4 (\Delta x)_t$, where $(\Delta x)_t =
\big( \sigma / \sqrt{2} \big) \sqrt{1 + (\hbar t / m \sigma^2)^2}$ is
the position uncertainty of the free-particle wave packet,
$\Psi_{\mathrm{free}}$, at time $t$. This choice guarantees that the
comparison between the AFM and DPM wave functions is performed on a
very broad spatial interval centered around the classically expected
position of the particle. For the parameter values specified above, we
have $(\Delta x)_t \simeq 30.1$~$\mu$m for $^7 \mathrm{Li}$,
22.2~$\mu$m for $^{23} \mathrm{Na}$, 21.5~$\mu$m for $^{41}
\mathrm{K}$, and 21.3~$\mu$m for $^{87} \mathrm{Rb}$.

\begin{figure}[ht]
\centering
\includegraphics[width=3.4in]{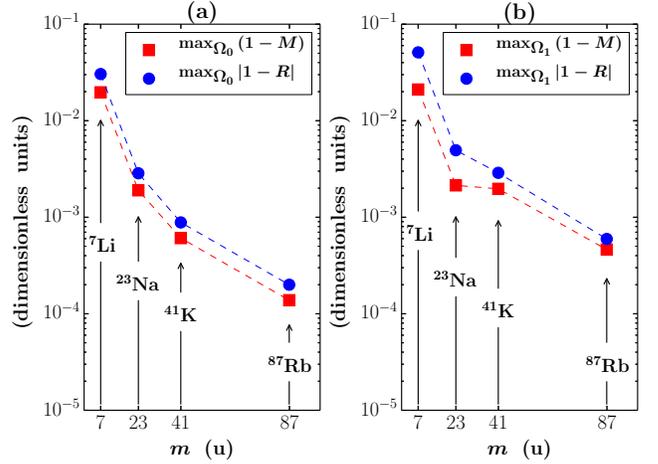}
\caption{(Color online) Maximal deviations of the fidelity $M$ (red
  squares) and probability ratio $R$ (blue circles) from 1 for four
  different alkali atoms. (a) The data represent the case of a
  time-independent barrier with $\Omega(\tau) = \Omega_0$; the maximal
  deviations are computed with respect to $\Omega_0$. (b) The data
  correspond to $\Omega(\tau) = \Omega_1 / \tau$; the maximal deviations are
  computed with respect to $\Omega_1$. See text for all parameter
  values.}
\label{FidRat_fig}
\end{figure}

Figure~\ref{FidRat_fig}(a) shows the values of the fidelity and
probability ratio deviations from 1, $(1-M)$ and $|1-R|$,
respectively, maximized over the wide range of barrier strengths $0
\leq \Omega_0 \leq 100 v_0$. We clearly see that the agreement between
$\Psi_{\afm}$ and $\Psi_{\dpm}$ significantly improves --
$\max_{\Omega_0} (1-M)$ and $\max_{\Omega_0} |1-R|$ decrease by over
two orders of magnitude -- as the atomic mass increases from
$m_{\mathrm{Li}}$ to $m_{\mathrm{Rb}}$. On a more practical side, in
the cases of potassium and rubidium, the wave functions predicted by
the AFM and DPM appear to be almost indistinguishable: already for
potassium, $M$ and $R$ deviate from 1 by less than 0.1\%.

\subsection{Algebraic barrier}
\label{One_over_t_BarSubsec}

As our first example of a time-dependent barrier, we consider the
scenario in which the atom-laser interaction is inversely proportional
to time, i.e., $\Omega(\tau) = \Omega_1 / \tau$. The DPM in this case
is exactly solvable. Indeed, introducing
$\psi_\pm = \psi_1 \pm \psi_2$ one rewrites Eq.~(\ref{SE-2}) as two
uncoupled Schr\"{o}dinger equations for a single-channel potential of
the form $\tau^{-1} \delta(x)$. The latter problem has been solved in
Ref.~\cite{SK88adiabaticity}. Using this solution, we obtain the
following expression for the full propagator of the DPM:
\begin{multline}
  \widehat{K}(x,x',\tau) = \widehat{K}_{0}(x-x',\tau) \\ -
  \frac{\Omega_1}{\Omega_1^2 + x'^2} K_{0}(\left\lvert x \right\rvert
  + \left\lvert x' \right\rvert,\tau)
  \begin{pmatrix} \Omega_1 & i
    |x'| \\[0.1cm] i |x'| & \Omega_1
  \end{pmatrix} \,.
\label{K_DPM_algTimeDep_expr}
\end{multline}

We compute the DPM wave function by substituting the first element of
the matrix propagator $\widehat{K}$,
\begin{equation}
  K_{11}(x,x',\tau) = K_0(x-x',\tau) - \frac{\Omega_1^2}{\Omega_1^2 +
    x'^2} K_0(|x|+|x'|,\tau) \,,
\end{equation}
into Eq.~(\ref{Psi_DPM_def}), and evaluating the resulting integral
numerically. The corresponding AFM wave function is obtained from
Eqs.~(\ref{Psi_AFM_general}) and (\ref{K_AFM_general}) by taking
$\chi(\tau) = \big[ 1 + ( \Omega_1 / v_0 \tau )^2\big]^{-1}$, in
accordance with Eq.~(\ref{chi_vs_Omega}). As in the time-independent
case of Sec.~\ref{TimeIndBarSubsec}, the fidelity and probability
ratio are computed by taking $x_A = x_t - 4 (\Delta x)_t$ and $x_B =
x_t + 4 (\Delta x)_t$.

Figure~\ref{FidRat_fig}(b) shows $\max_{\Omega_1} (1-M)$ (red squares)
and $\max_{\Omega_1} |1-R|$ (blue circles) corresponding to the range
$0 \le \Omega_1 \le 100 v_0 t_c$. Once again, we observe the agreement
between the predictions of the AFM and DPM improve as the mass of the
atom increases. In particular, in the case of rubidium, the fidelity
and probability ratio deviate from 1 by less than 0.1\%.

\subsection{Exponential barriers}
\label{Exponential_BarSubsec}

\begin{figure*}[ht]
\centering
\includegraphics[width=4.5in]{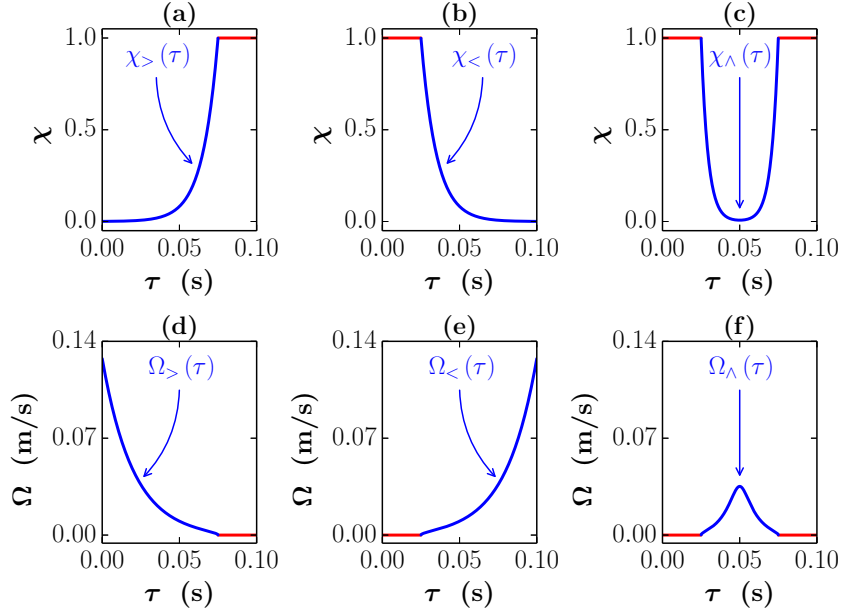}
\caption{(Color online) Aperture function $\chi(\tau)$ and the
  corresponding atom-laser interaction strength $\Omega(\tau)$ in
  three different scenarios. (a) $\chi(\tau) = \min \big\{
  \chi_>(\tau), 1 \big\}$, with $\chi_> (\tau) \equiv \exp[ \gamma
    (\tau - 3 t_c / 2) ]$ and $\gamma = 100$~s$^{-1}$, (b) $\chi(\tau)
  = \min \big\{ \chi_<(\tau), 1 \big\}$, with $\chi_< (\tau) \equiv
  \exp[ \gamma (\tau - t_c / 2) ]$ and $\gamma = -100$~s$^{-1}$, (c)
  $\chi(\tau) = \min \big\{ \chi_{\wedge}(\tau), 1 \big\}$, with
  $\chi_{\wedge} (\tau) \equiv \cosh[ \gamma (\tau - t_c) ] / \cosh
  (\gamma t_c / 2)$ and $\gamma = 225$~s$^{-1}$, (d) $\Omega(\tau) =
  \Omega_>(\tau) \equiv v_0 \sqrt{1 / \chi_>(\tau) - 1}$ for $\tau < 3
  t_c / 2$, and $\Omega(\tau) = 0$ for $\tau \ge 3 t_c / 2$, (e)
  $\Omega(\tau) = 0$ for $\tau \le t_c / 2$, and $\Omega(\tau) =
  \Omega_<(\tau) \equiv v_0 \sqrt{1 / \chi_<(\tau) - 1}$ for $\tau >
  t_c / 2$, and (f) $\Omega(\tau) = \Omega_{\wedge}(\tau) \equiv v_0
  \sqrt{1 / \chi_{\wedge}(\tau) - 1}$ for $|\tau - t_c| < t_c / 2$,
  and $\Omega(\tau) = 0$ for $|\tau - t_c| \ge t_c / 2$.}
\label{exponential_type_barriers_fig}
\end{figure*}

We now consider absorbing barriers whose aperture function
$\chi(\tau)$ exhibits exponential dependence on time on some
intervals. As recently shown in Ref.~\cite{Gou15Manipulating}, such
barriers can be efficiently used to manipulate, e.g., shift, split, or
squeeze, the spatial wave function of the transmitted particle. Here,
we consider three different scenarios defined by the aperture
functions presented in
Figs.~\ref{exponential_type_barriers_fig}(a)--\ref{exponential_type_barriers_fig}(c).
The corresponding atom-laser interaction strengths are computed using
Eq.~(\ref{chi_vs_Omega}) and shown in
Figs.~\ref{exponential_type_barriers_fig}(d)--\ref{exponential_type_barriers_fig}(f). The
AFM wave function is calculated by evaluating the integrals in
Eqs.~(\ref{Psi_AFM_general}) and (\ref{K_AFM_general}). The DPM wave
function is obtained by solving the time-dependent Schr\"{o}dinger
equation, Eqs.~(\ref{SE-2})--(\ref{V(x,t)}), numerically, using a
Crank-Nicolson algorithm \cite{CN47practical}.

\begin{figure*}[ht]
\centering
\includegraphics[width=6in]{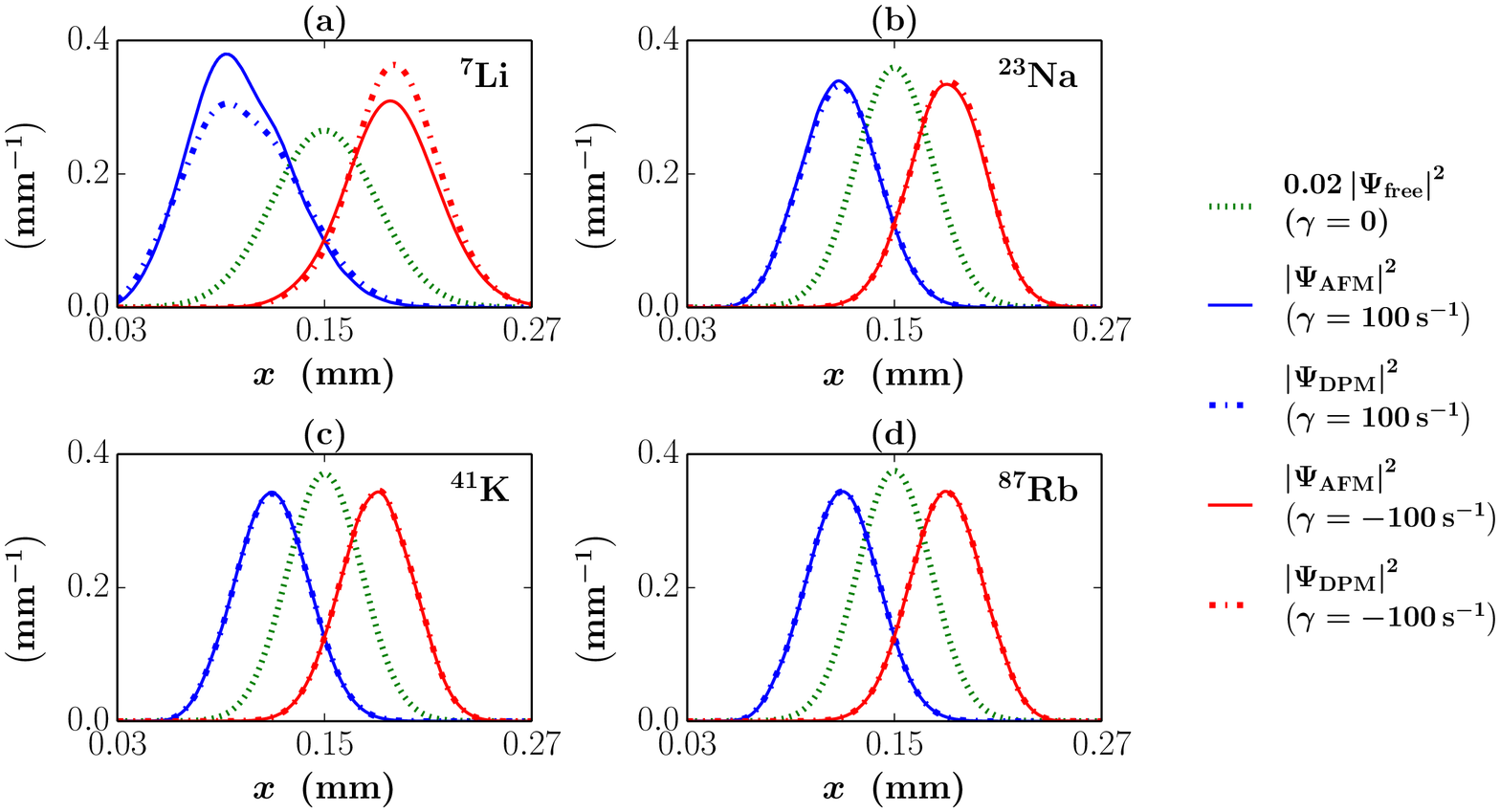}
\caption{(Color online) Probability densities $|\Psi_{\afm}|^2$ (solid
  curves) and $|\Psi_{\dpm}|^2$ (dash-dotted curves) for (a)
  $^7 \mathrm{Li}$, (b) $^{23} \mathrm{Na}$, (c) $^{41} \mathrm{K}$
  and (d) $^{87} \mathrm{Rb}$ as functions of the position $x$. Blue
  curves correspond to the barrier specified in
  Figs.~\ref{exponential_type_barriers_fig}(a) and
  \ref{exponential_type_barriers_fig}(d). Red curves correspond to the
  barrier specified in Figs.~\ref{exponential_type_barriers_fig}(b)
  and \ref{exponential_type_barriers_fig}(e). The dotted green curve
  represents the scaled probability density of the corresponding
  free-particle wave packet.}
\label{Shifting_ProbaDens_fig}
\end{figure*}

Figure~\ref{Shifting_ProbaDens_fig} shows the probability densities
$|\Psi_{\afm}(x,t)|^2$ (solid curves) and $|\Psi_{\dpm}(x,t)|^2$
(dash-dotted curves) for all four atoms as functions of $x$. Blue
curves correspond to the barrier defined by $\chi(\tau) = \min \big\{
\chi_>(\tau), 1 \big\}$, with $\chi_> (\tau) = \exp[ \gamma (\tau - 3
  t_c / 2) ]$ and $\gamma = 100$~s$^{-1}$ [see
  Figs.~\ref{exponential_type_barriers_fig}(a) and
  \ref{exponential_type_barriers_fig}(d)]. Red curves correspond to
the barrier defined by $\chi(\tau) = \min \big\{ \chi_<(\tau), 1
\big\}$, with $\chi_< (\tau) = \exp[ \gamma (\tau - t_c / 2) ]$ and
$\gamma = -100$~s$^{-1}$ [see
  Figs.~\ref{exponential_type_barriers_fig}(b) and
  \ref{exponential_type_barriers_fig}(e)]. The dotted green curve
represents the scaled probability density of the free-particle
Gaussian wave packet, $0.02 \, |\Psi_{\mathrm{free}}(x,t)|^2$, that
would be observed in the absence of a barrier, i.e., for $\gamma =
0$. The effect of the barrier is to shift the transmitted wave packet
by the distance $\Delta \simeq -\gamma \sigma^2 / v_0$ with respect to
the position of the freely evolved Gaussian~\cite{Gou15Manipulating}.

\begin{figure*}[ht]
\centering
\includegraphics[width=6in]{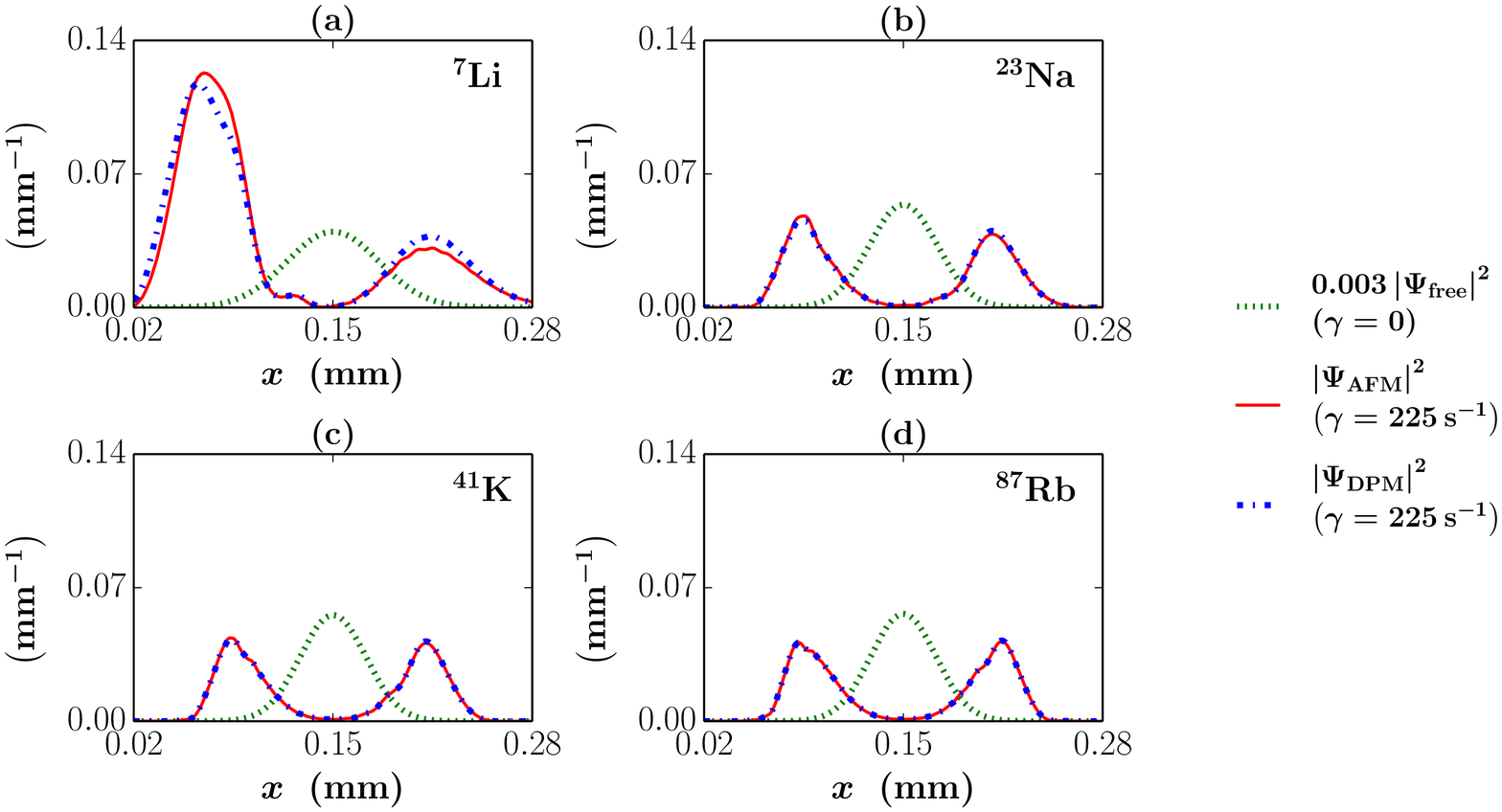}
\caption{(Color online) Probability densities $|\Psi_{\afm}|^2$ (solid
  red curve) and $|\Psi_{\dpm}|^2$ (dash-dotted blue curve) for (a)
  $^7 \mathrm{Li}$, (b) $^{23} \mathrm{Na}$, (c) $^{41} \mathrm{K}$
  and (d) $^{87} \mathrm{Rb}$ as functions of the position $x$. The
  curves correspond to the barrier specified in
  Figs.~\ref{exponential_type_barriers_fig}(c) and
  \ref{exponential_type_barriers_fig}(f). The dotted green curve
  represents the scaled probability density of the corresponding
  free-particle wave packet.}
\label{Splitting_ProbaDens_fig}
\end{figure*}

Figure~\ref{Splitting_ProbaDens_fig} shows the probability densities
predicted by the AFM (solid red curve) and DPM (dash-dotted blue
curve) for the barrier defined by $\chi(\tau) = \min \big\{
\chi_{\wedge}(\tau), 1 \big\}$, with $\chi_{\wedge} (\tau) \equiv
\cosh[ \gamma (\tau - t_c) ] / \cosh (\gamma t_c / 2)$ and $\gamma =
225$~s$^{-1}$ [see Fig.~~\ref{exponential_type_barriers_fig}(c) and
  \ref{exponential_type_barriers_fig}(f)]. For reference, the dotted
green curve shows the scaled probability density of the free-particle
Gaussian wave packet, $0.003 \, |\Psi_{\mathrm{free}}(x,t)|^2$. The
effect of the barrier is to spatially split the transmitted wave
packet in two as compared to the freely evolved Gaussian wave packet.

As in the examples considered in Secs.~\ref{TimeIndBarSubsec} and
\ref{One_over_t_BarSubsec}, we observe in
Figs.~\ref{Shifting_ProbaDens_fig} and \ref{Splitting_ProbaDens_fig}
that the agreement between the AFM and DPM quickly improves as the
atomic mass is increased. More quantitatively, the values of $(1-M)$
and $|1-R|$ decrease by approximately two orders of magnitude as we go
from the case of lithium to that of rubidium.

\subsection{Moshinsky shutter}
\label{Moshinsky_BarSubsec}

We conclude this section by considering the case of the Moshinsky
shutter, i.e. an instantaneously opening barrier with the aperture
function defined by Eq.~(\ref{defAF_Moshi}). As in the previous
examples, $\Psi_{\afm}$ is calculated by using
Eqs.~(\ref{Psi_AFM_general}) and (\ref{K_AFM_general}), whereas
$\Psi_{\dpm}$ is obtained from Eq.~(\ref{Psi_DPM_def}) with the
propagator $K_{11}$ given by Eq.~(\ref{K_11_Moshinsky}).

\begin{figure*}[ht]
\centering
\includegraphics[width=6in]{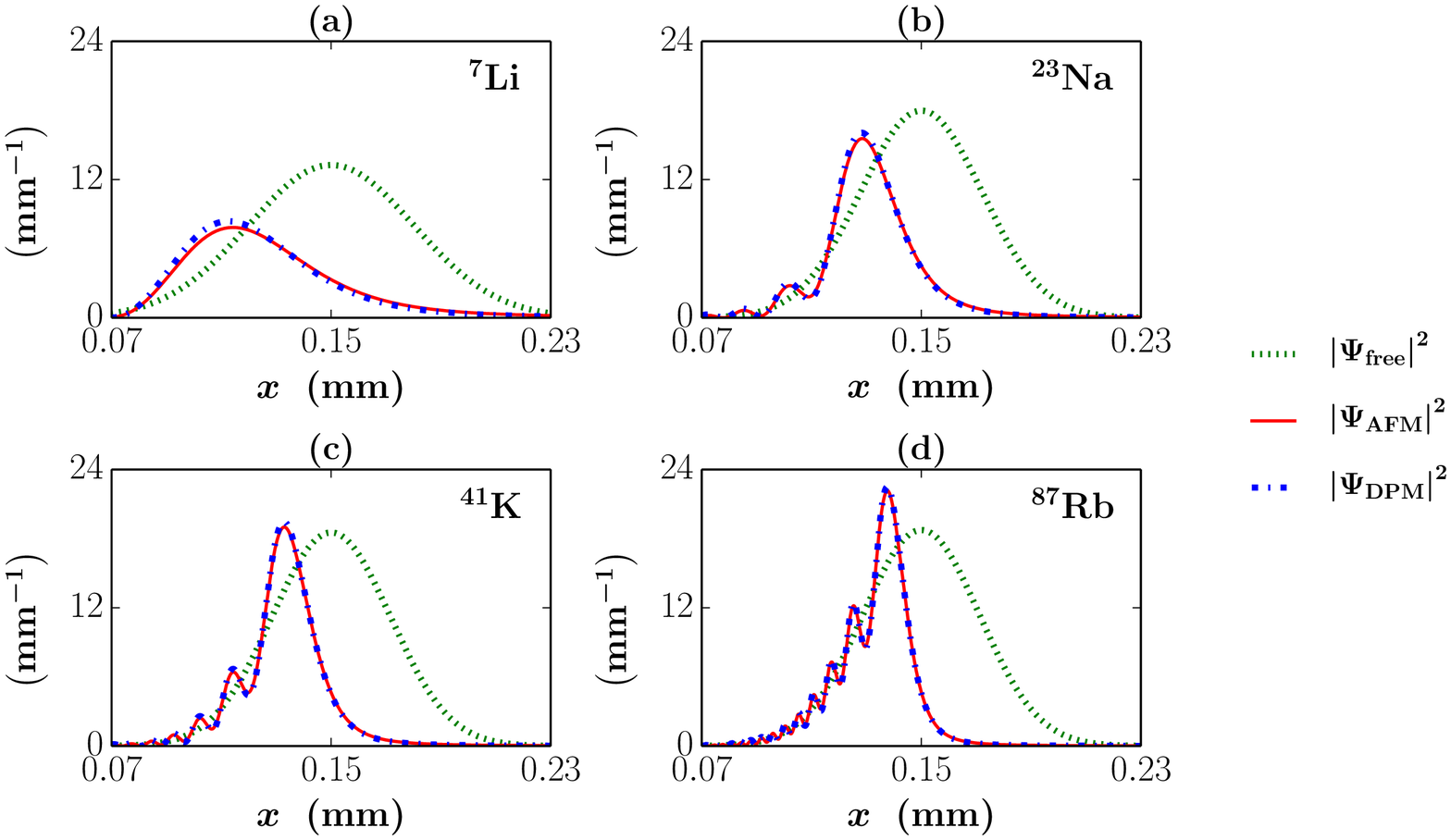}
\caption{(Color online) Probability densities $|\Psi_{\afm}(x,t)|^2$
  (solid red curve) and $|\Psi_{\dpm}(x,t)|^2$ (dash-dotted blue
  curve) for (a) $^7 \mathrm{Li}$, (b) $^{23} \mathrm{Na}$, (c) $^{41}
  \mathrm{K}$ and (d) $^{87} \mathrm{Rb}$ evaluated for the case of
  the Moshinsky shutter, Eq.~(\ref{defAF_Moshi}). The dotted green
  curve represents the probability density of the corresponding
  free-particle wave packet.}
\label{Moshinsky_ProbaDens_fig}
\end{figure*}

Figure~\ref{Moshinsky_ProbaDens_fig} shows the probability densities
$|\Psi_{\afm}|^2$ (solid red curve), $|\Psi_{\dpm}|^2$ (dash-dotted
blue curve), and $|\Psi_{\mathrm{free}}|^2$ (dotted green curve) as
functions of the position $x$. The Moshinsky barrier induces
oscillations of the probability density of the transmitted wave packet
as compared to that of a freely evolved Gaussian. These oscillations
get more and more pronounced as the mass of the atom increases.

In accord with all other examples of this section, we find better
agreement between the predictions of the AFM and DPM for heavier (more
semiclassical) atoms. In particular, the deviation of the fidelity
from 1, $(1-M)$, decreases by approximately 14 times as we go from the
lithium to the rubidium system; the deviation of the probability
ratio, $|1-R|$, decreases by approximately 17 times.


\section{Summary and conclusion}
\label{sec:conclusion}

In this paper we compared two different theoretical descriptions of
the motion of a quantum particle, e.g., an atom, through a
time-dependent absorbing point-like barrier, e.g., a sheet of laser
light with a time-dependent intensity. The first one, the aperture
function model (AFM), represents the barrier by a set of
time-dependent discontinuous matching conditions of Kottler type,
Eqs.~(\ref{BC-1}) and (\ref{BC-2}). The key ingredient of the model is
a time-dependent transmission amplitude $\chi(\tau)$, entering the
matching conditions. The main advantage of the AFM is that it allows
for an explicit integral expression of the quantum propagator,
Eq.~(\ref{K_AFM_general}). The second description, the delta potential
model (DPM), represents the barrier by an off-diagonal
$\delta$-potential with a time-dependent amplitude $\Omega(\tau)$,
Eq.~(\ref{V(x,t)}). The DPM is a time-dependent generalization of the
stationary atom-laser interaction model addressed in
Ref.~\cite{CM06Exact}. To date, only few exactly solvable cases of the
DPM are known, examples being the systems characterized by
$\Omega(\tau) = \mathrm{constant}$ and
$\Omega(\tau) = \mathrm{constant}/\tau$. In general, the DPM has to be
solved by numerically integrating the time-dependent Schr\"{o}dinger
equation, Eqs.~(\ref{SE-2})--(\ref{V(x,t)}).

Here, we have found a way of mapping the two models onto one another
in the semiclassical regime. More concretely, we have shown that, in
the transmission region, the wave functions predicted by the AFM and
DPM are in good agreement with each other provided that the aperture
function $\chi$ and the amplitude $\Omega$ are related through
Eq.~(\ref{chi_vs_Omega}). The agreement improves as the mass of the
moving particle increases, i.e. as the system becomes more
semiclassical. Our conclusion is based on both asymptotic analytical
calculations, presented in Sec.~\ref{sec:semiclassics}, and a detailed
numerical investigation of four particle-barrier systems with
experimentally realistic parameters, reported in
Sec.~\ref{sec:numerics}.

Time-dependent absorbing barriers may be realized in atom-optics
laboratory experiments. Recently, such barriers were identified as a
promising tool for engineering and reshaping (e.g., shifting,
splitting, squeezing, and cooling) of atomic wave packets
\cite{Gou15Manipulating}. The main practical value of the present
study is that we have extended the range of theoretical tools
appropriate for investigating and making quantitative predictions
related to light-based manipulation of atomic wave functions. In
future, it would be interesting to construct and explore other
representations of time-dependent point-like absorbing barriers, some
promising candidates being the imaginary $\delta$-potential
\cite{MSTotal}, and generalized point interactions of the form $c_1
\delta(x) + c_2 \delta'(x)$
\cite{AGHH05Solvable,CLMM14Distributional}.


\begin{acknowledgments}
  The authors are thankful to Adolfo del Campo for useful
  suggestions. A.G. acknowledges the financial support of EPSRC under
  Grant No. EP/K024116/1. M. Beau acknowledges funding support from ESF (POLATOM-5052).
\end{acknowledgments}

\appendix

\section{Derivation of Eq.~\eqref{detHessian}}
\label{appA}

We start by rewriting the Hessian, given by Eq.~\eqref{HessianElem},
in the matrix form
\begin{equation}
  \mathcal{H} = 2 \left( \sum\limits_{k=1}^{n+1} \lvert \epsilon_{k} -
  \epsilon_{k-1} \rvert \right)^3 \mathcal{A} \,,
\label{HwithA}
\end{equation}
where
\begin{align}
  A = \begin{pmatrix} b_1 & c_1 & 0  & \cdots & \cdots & 0
    \\ a_2 & \ddots & \ddots  & \ddots & \, & \vdots \\ 0 & \ddots
    & \ddots  & \ddots & \ddots & \vdots \\ \vdots & \ddots & \ddots
    & \ddots & \ddots & 0 \\ \vdots & \, & \ddots &
    \ddots & \ddots & c_{n-1} \\ 0 & \cdots & \cdots & 0 &
    a_n & b_n
\end{pmatrix}
\label{A_def}
\end{align}
with
\begin{subequations}
\label{A_elem}
\begin{equation}
  a_j = - \frac{1}{\lvert \epsilon_{j} - \epsilon_{j-1} \rvert} \,,
  \quad 2 \leq j \leq n \,,
\label{A_elem_subDiag}
\end{equation}
\begin{equation}
  b_j = \frac{1}{\lvert \epsilon_{j} - \epsilon_{j-1} \rvert} +
  \frac{1}{\lvert \epsilon_{j+1} - \epsilon_{j} \rvert} \,, \quad 1
  \leq j \leq n \,,
\label{A_elem_Diag}
\end{equation}
and
\begin{equation}
  c_j = - \frac{1}{\lvert \epsilon_{j+1} - \epsilon_{j} \rvert} \,,
  \quad 1 \leq j \leq n-1 \,.
\label{A_elem_upDiag}
\end{equation}
\end{subequations}
It follows immediately from Eq.~\eqref{HwithA} that
\begin{equation}
  \det(\mathcal{H}) = 2^n \left( \sum\limits_{k=1}^{n+1} \lvert
  \epsilon_{k} - \epsilon_{k-1} \rvert \right)^{3n} \det(\mathcal{A})
  \,.
\label{detH_detA}
\end{equation}

In order to find $\det(\mathcal{A})$ we use an LU decomposition of the
matrix $\mathcal{A}$. That is, we express $\mathcal{A}$ as
\begin{equation}
  \mathcal{A} = \mathcal{L} \, \mathcal{U} \, \text{,}
\label{A_LU}
\end{equation}
where
\begin{align}
  \mathcal{L} = \begin{pmatrix} 1 & 0 & 0  & \cdots & \cdots &
    0 \\ L_2 & \ddots & \ddots & \ddots & \, & \vdots \\ 0 &
    \ddots & \ddots & \ddots & \ddots & \vdots \\ \vdots & \ddots &
    \ddots & \ddots & \ddots & 0 \\ \vdots & \, & \ddots
    & \ddots & \ddots & 0 \\ 0 & \cdots & \cdots & 0 & L_n &
    1
\end{pmatrix}
\label{L_matrix_def}
\end{align}
and
\begin{align}
  \mathcal{U} = \begin{pmatrix} U_1 & c_1 & 0 & \cdots &
    \cdots & 0 \\ 0 & \ddots & \ddots & \ddots & \, & \vdots \\ 0
    & \ddots & \ddots & \ddots & \ddots & \vdots \\ \vdots & \ddots
    & \ddots & \ddots & \ddots & 0 \\ \vdots & \, & \ddots & 
    \ddots & \ddots & c_{n-1} \\ 0 & \cdots & \cdots
    & 0 & 0 & U_n
\end{pmatrix} \,.
\label{U_matrix_def}
\end{align}
Substituting Eqs.~\eqref{A_def}, \eqref{L_matrix_def}, and
\eqref{U_matrix_def} into Eq.~\eqref{A_LU}, we see that the $(2n - 1)$
matrix elements $L_j$ and $U_j$ must satisfy the following $(2n - 1)$
equations:
\begin{equation}
  \left\{
  \begin{array}{ll}
    U_1 = b_1 & \\ L_{j+1} U_{j} = a_{j+1} & \,, \quad 1 \leq j \leq
    n-1 \\ L_{j+1} c_{j} + U_{j+1} = b_{j+1} & \,, \quad 1 \leq j \leq
    n-1
  \end{array}
  \right. \,.
\label{LU_elemCond}
\end{equation}
Solving this system of equations, we find
\begin{equation}
  L_j = - \, \frac{\sum\limits_{k=1}^{j-1} \lvert \epsilon_{k} -
    \epsilon_{k-1} \rvert}{\sum\limits_{k=1}^{j} \lvert \epsilon_{k} -
    \epsilon_{k-1} \rvert}
\label{L_elem_expr}
\end{equation}
for all $2 \leq j \leq n$, and
\begin{equation}
  U_j = \frac{\sum\limits_{k=1}^{j+1} \lvert \epsilon_{k} -
    \epsilon_{k-1} \rvert}{\lvert \epsilon_{j+1} - \epsilon_{j} \rvert
    \sum\limits_{k=1}^{j} \lvert \epsilon_{k} - \epsilon_{k-1} \rvert}
\label{U_elem_expr}
\end{equation}
for all $1 \leq j \leq n$.

It is now straightforward to compute the determinant of
$\mathcal{A}$. In view of Eqs.~\eqref{A_LU}, \eqref{L_matrix_def} and
\eqref{U_matrix_def}, we have
\begin{equation}
  \mathrm{det}(\mathcal{A}) = \mathrm{det}(\mathcal{L})
  \mathrm{det}(\mathcal{U}) = \prod\limits_{j=1}^{n} U_j \,,
\label{detA_detU}
\end{equation}
and thus, using Eq.~\eqref{U_elem_expr},
\begin{equation}
  \det(\mathcal{A}) = \frac{\sum\limits_{k=1}^{n+1} \lvert
    \epsilon_{k} - \epsilon_{k-1} \rvert}{\prod\limits_{k=1}^{n+1}
    \lvert \epsilon_{k} - \epsilon_{k-1} \rvert} \,.
\label{detA_expr}
\end{equation}
Finally, combining Eq.~\eqref{detH_detA} and \eqref{detA_expr}, we
arrive at the final result, Eq.~\eqref{detHessian}.

%

\end{document}